\documentclass[12pt]{article}
\usepackage{amsmath}
\usepackage{graphicx,psfrag,epsf}
\usepackage{enumerate}
\usepackage{url} 
\usepackage{natbib}
\setcitestyle{aysep={}} 
\usepackage{amsthm}
\usepackage[font=small, labelsep=period, labelfont=bf, justification=justified, singlelinecheck=false]{caption} 
\usepackage{setspace}
\newtheorem{theorem}{Theorem}
\theoremstyle{definition}
\newtheorem*{definition}{Definition}
\usepackage[top    = 1.1in, bottom = 1.1in, left   = 1in, right  = 1in]{geometry}
\renewcommand{\baselinestretch}{2}


\usepackage{color}
\definecolor{red}{rgb}{1,0,0}
\definecolor{blue}{rgb}{0,0,1}
\definecolor{green}{rgb}{0,0.6,0.4}

\newcommand{\blind}{1}

\begin{document}

\pagenumbering{gobble} 

\def\spacingset#1{\renewcommand{\baselinestretch}%
{#1}\small\normalsize} \spacingset{1}

\if1\blind
{
  \title{\bf \large Testing a Large Number of Composite Null Hypotheses Using Conditionally Symmetric Multidimensional Gaussian Mixtures in Genome-Wide Studies}
 \author{Ryan Sun, Zachary McCaw, Xihong Lin
   \thanks{Ryan Sun is Assistant Professor  in the Department of Biostatistics at MD Anderson Cancer Center ({\em RSun3@mdanderson.org}). Zachary McCaw is Senior Machine Learning Scientist at Insitro ({\em zrmacc@gmail.com}). Xihong Lin is Professor of Biostatistics at Harvard T.H. Chan School of Public Health and Professor of Statistics at Harvard University ({\em xlin@hsph.harvard.edu}). This work  was supported by grants R35-CA197449, U19-CA203654  and U01-CA209414 from the National Cancer Institute, U01-HG009088 and U01-HG012064 from the National Human Genome Research Institute, and R01-HL113338 from the National Heart, Lung, and Blood Institute.}}

  \maketitle
} \fi

\if0\blind
{
  \bigskip
  \bigskip
  \bigskip
  \begin{center}
    {\LARGE\bf Testing a Large Number of Composite Null Hypotheses Using Conditionally Symmetric Multidimensional Gaussian Mixtures in Genome-Wide Studies}
\end{center}
  \medskip
} \fi


\begin{abstract}
Causal mediation analysis, pleiotropy analysis, and replication analysis are three highly popular
genetic study designs.
Although these analyses address different scientific questions, the underlying inference
problems all involve large-scale testing of composite null hypotheses.
The goal is to determine whether all null hypotheses - as opposed to at least one - in a set
of individual tests should simultaneously be rejected. 
Various recent methodology has been proposed for the aforementioned situations, and an appealing empirical Bayes strategy is to apply the popular two-group model,
calculating local false discovery rates (lfdr) for each set of hypotheses. 
However, such a strategy is difficult due to the need for multivariate density estimation.
Furthermore, the multiple testing rules for the empirical Bayes lfdr approach and conventional frequentist z-statistics can disagree, 
which is troubling for a field that ubiquitously utilizes summary statistics. 
This work proposes a framework to unify two-group testing in genetic association composite null settings,
the conditionally symmetric multidimensional Gaussian mixture model (csmGmm).
The csmGmm is shown to demonstrate more robust operating characteristics than recently-proposed alternatives.
Crucially, the csmGmm also offers strong interpretability guarantees by harmonizing lfdr and z-statistic testing rules.
We extend the base csmGmm to cover each of the mediation, pleiotropy, and replication settings, and we
prove that the lfdr z-statistic agreement holds in each situation.
We apply the model to a collection of translational lung cancer genetic association studies that motivated this work. 

\end{abstract}

\noindent%
\vfill

\newpage
\spacingset{1.45} 

\pagenumbering{arabic}
\baselineskip=24pt

\section{Introduction}
\label{sec:intro}

Three of the most popular genetic study designs in the post genome-wide association study (GWAS) era include
mediation analysis, pleiotropy analysis, and replication analysis \citep{visscher201710}.
Reasons for their broad appeal and utilization across numerous phenotypes include an ability to produce
translationally-relevant insights and applicability
across multiple forms of omic data \citep{dai2022multiple, pettit2021shared, byun2022cross}.
However, all three designs are also challenged by a similar underlying statistical inference problem,
the difficulty of large-scale composite null hypothesis testing across the genome.
Motivating examples commonly arise in translational lung cancer genetics research, 
where it has been of much interest to identify single nucleotide polymorphisms (SNPs)
with mediated, pleiotropic, and replicated effects \citep{mckay2017large, pettit2021shared, byun2022cross}.

In the aforementioned genetic contexts, a composite null hypothesis arises when attempting to
test whether a set of two or more individual null hypotheses, e.g. (i) $H_{0,1}: \alpha = 0$ and (ii) $H_{0,2}: \beta = 0$,
are all false \citep{barfield2017testing}.
The composite null then has the form $H^{c}_{0}: \alpha = 0 \cup \beta=0$  \citep{dai2022multiple, liu2022large}.
For example, in lung cancer, mediation analysis is of great interest because many SNPs are believed to perturb risk of disease through
effects that are mediated by gene expression  \citep{mckay2017large}.
To identify such an effect, it is customary to test (i) the association between a SNP and gene expression ($\alpha$) and (ii) 
the association between that same gene expression and lung cancer ($\beta$).
There is a mediated effect if both (i) and (ii) exist at the same time \citep{huang2019genome},  and thus $H^{c}_{0}$ is the null of interest.

Much recent interest has also focused on identifying pleiotropic variants affecting risk of both lung cancer
and related diseases \citep{byun2021shared}.
Motivated by secondary clinical trial results demonstrating the efficacy of cardiovascular drugs
for treating lung cancer, researchers have searched for SNPs that are simultaneously associated with 
both (i) lung cancer ($\alpha$) and (ii) another outcome ($\beta$) \citep{pettit2021shared}.
The pleiotropy setting is more complex than mediation in that 
test statistics for these parameters may be correlated if outcomes are measured in the same cohort.
Also, more parameters may be involved, e.g. in pleiotropy analysis of three outcomes.
Replication studies are perpetually necessary to confirm novel findings and are
 similar to pleiotropy inference in that there may be more than two parameters.
However, replication analysis is unique in that effects 
are only considered replicated if they point in the same direction, e.g. replicated effects of a SNP on lung cancer parameterized by (i) $\alpha$ in one cohort and (ii) $\beta$ in another should possess the same sign.
 
It is well-known that traditional inference approaches such as a Wald or likelihood ratio test do not
perform well in the applications of interest \citep{barfield2017testing, huang2019genome,
dai2022multiple}.
Briefly, $H_{0}^{c}$ is true if $\alpha=\beta=0$ (case 0)
and also if $\alpha = 0, \beta \neq 0$ (case 1) or $\alpha \neq 0, \beta = 0$ (case 2), and thus $(\alpha,\beta)$ can take an infinite number of values under the composite null.
There is no longer a single reference distribution for calculating p-values, unlike 
the global null $H_{0}^{g}: \alpha = 0 \cap \beta=0$.
Recently, some authors \citep{dai2022multiple, liu2022large} have proposed strategies
that involve estimating the probability of cases 0-2 and then applying these
estimates to construct improved tests.
While effective results have been observed in certain applications, such approaches still have limitations.
Less optimal performance can be observed when non-zero effect sizes are larger or more frequent
\citep{dai2022multiple, liu2022large}.
Furthermore, the strategies do not easily generalize to more complicated testing situations such as 
composite null sets with more than two individual nulls, settings with correlated outcomes, or applications that consider
effect direction.

Another possible testing approach is to adopt an empirical Bayes strategy
using the popular two-group model \citep{efron2002empirical, efron2008microarrays},
which has been proposed for replication analysis \citep{heller2014replicability}.
Composite null hypothesis sets can be declared significant based on local false discovery rates (lfdr), 
which enjoy desirable properties in large-scale multiple testing problems \citep{sun2007oracle}.
However, a major obstacle in adopting the lfdr approach is the need to perform multivariate 
density estimation, which is quite challenging \citep{efron2008microarrays}.
While various multivariate density models exist \citep{heller2014replicability, wang2020imix}, they often possess unclear 
operating characteristics or may not be flexible enough for each of the aforementioned composite null settings,
which all present unique challenges.

Additionally, one critical drawback of the empirical Bayes lfdr approach is that it can contradict conventional
frequentist z-statistics when defining a multiple testing rule.
For example, in our mediation data analysis, we show how the lfdr-value for a set of z-statistics $(2.85, -4.40)$ can be larger, 
i.e. less significant, than the lfdr-value for the z-statistics $(2.65, -4.14)$.
As a result, the latter set is declared significant while the former is not.
This behavior conflicts with both intuition and traditional multiple testing
rules based on the frequentist z-statistics, where $(2.85, -4.40)$ would be rejected because it is more extreme and falls in the same
directions as $(2.65, -4.14)$.
Such a discrepancy is generally very confusing and uninterpretable for practitioners, especially given the ubiquity of summary statistics
in genetics research \citep{visscher201710}.
The technical explanation relies on the density estimation required to calculate lfdr-values, 
which motivates the question of whether any estimation approach can perform well while preventing such 
hard-to-interpret results.

This work addresses the above challenges with three main contributions.
First, we propose a new density model for empirical Bayes analysis of composite null hypotheses,
the conditionally symmetric multidimensional
Gaussian mixture model (csmGmm), which unifies multiple common genetic analyses under a single framework.
Broadly, the main theme of the csmGmm is to place the right type and right amount of constraints on the density model such 
that it offers good performance while remaining flexible enough for each of the three main genetic applications.
Second, we prove that desirable interpretability guarantees hold across each of the mediation, pleiotropy, and replication settings,
as the model enforces harmonization of empirical Bayes lfdr and frequentist z-statistic testing rules.
Finally, we demonstrate that the csmGmm provides more robust operating characteristics than existing alternatives
over a wide range of realistic composite null settings.
Specifically, extensive simulations demonstrate that the csmGmm offers both stringent control of error rates
and superior power profiles over a wider range of settings than state-of-the-art methodology \citep{heller2014replicability, dai2022multiple, liu2022large}.

We demonstrate the flexibility and wide applicability of the csmGmm by performing
a varied suite of lung cancer analyses that motivated this work.
Our main investigation assesses whether top lung cancer GWAS SNPs are
perturbing disease risk through effects mediated by gene expression.
We also search for pleiotropic SNPs that are associated with the three closely-related 
traits lung cancer, coronary artery disease (CAD), and body mass index (BMI).
These outcomes are often studied together because of their hypothesized common etiology
and potential to act as mutual risk factors.
Lastly, we perform within-lung cancer and within-CAD replication of SNPs found significant in the  
pleiotropy analysis.

The remainder of the manuscript is organized as follows.
In Section \ref{sec:individualnull} we introduce the mediation, pleiotropy, and replication settings and describe
how they fall into the composite null framework.
In Section  \ref{sec:csmgmm} we propose the csmGmm model and its extensions, and we provide theorems that
ensure the interpretability of the model results.
Section \ref{sec:empbayes} provides details on empirical Bayes inference using the csmGmm.
Section  \ref{sec:simulation} describes simulations and comparisons to existing
methods, and Section \ref{sec:application} demonstrates three applications across varied datasets.
We conclude with a discussion in Section \ref{sec:discussion}.


\section{Models for Mediation, Pleiotropy, and Replication}
\label{sec:individualnull}

\subsection{Mediation Study}
 \label{ss:mediationmodel}
 
In a genetic mediation study, suppose for each of $i=1,...,n$ subjects we observe a single scalar genotype $G_{i}$, 
an outcome $Y_{i}$, the $J$ mediators $M_{ij}$ $(j=1,...,J)$, and a vector of
$p$ additional covariates $\mathbf{X}_{i} = (X_{i1},...,X_{ip})^{T}$. 
In lung cancer examples, $G_{i}$ is a SNP of interest, $Y_{i}$ is the binary lung cancer status, and the $M_{ij}$ are 
lung tissue gene expression values for $J$ genes across the genome. 
We want to perform mediation analysis for each mediator separately by fitting the $J$ pairs of regression models \citep{dai2022multiple, liu2022large}
\begin{align}
 M_{ij} = & \alpha_{j0} + \boldsymbol{\alpha}_{jX}^{T} \mathbf{X}_{i} + \alpha_{j}G_{i}, \label{eq:med1} \\
 \mbox{logit}(\mu_{ij})  = & \beta_{j0} + \boldsymbol{\beta}_{jX}^{T} \mathbf{X}_{i} +  \beta_{jG}G_{i} + \beta_{j} M_{ij}, \label{eq:med2}
 \end{align}
 where $\mu_{ij} = E(Y_{i} | \mathbf{X}_{i}, M_{ij}, G_{i})$. 
Let $Z_{j1}$ be the result of a score test for $H_{0,j,1}:\alpha_{j} = 0$, and let $Z_{j2}$ be the result of a score test for $H_{0,j,2}:\beta_{j} = 0$.
It can be shown that $Z_{j1}$ and $Z_{j2}$ are independent \citep{liu2022large}. 
Under some common assumptions, the SNP $G_{i}$ possesses an effect on lung cancer mediated through expression of gene $j$ 
 if we reject the composite null $H^{med}_{0,j}: \alpha_{j} = 0 \cup \beta_{j}=0$ \citep{huang2019genome}. 
This basic two-regression framework can be adapted in many ways, for instance 
Equations (\ref{eq:med1}) and (\ref{eq:med2})
could be applied in separate datasets if the data are not all available in one package.
However, it is clear to see that the form of the composite null $H^{med}_{0,j}$ would be the same.

\subsection{Pleiotropy and Replication}
 \label{ss:pleiotropymodel}
 
In a pleiotropy study, suppose that we have two GWAS datasets A and B.
For instance, A may be a GWAS for lung cancer in the International Lung Cancer Consortium (ILCCO) dataset \citep{mckay2017large},
 and B may be a GWAS for coronary artery disease (CAD) in the UK Biobank (UKB) dataset \citep{bycroft2018uk}.
In cohort A, suppose that for each of $i=1,...,n_{A}$ subjects we observe 
genotype data $\mathbf{G}_{i}^{A} = (G_{i1}^{A},...,G_{iJ}^{A})^{T}$ for $J$ total
SNPs as well as the outcome $Y_{i}^{A}$ and a vector of $p$ additional covariates
$\mathbf{X}_{i}^{A} = (X_{i1}^{A},...,X_{ip}^{A})^{T}$. 
Let $\mathbf{G}_{i'}^{B}$,  $Y_{i'}^{B}$, and $\mathbf{X}_{i'}^{B}$ be defined similarly for the
$i'=1,...,n_{B}$ subjects in cohort B, where the $j$th genotype always indexes the same SNP in both datasets.
A pleiotropy analysis fits
\begin{align}
\mbox{logit}(\mu_{i}^{A}) = \alpha_{j0} + \boldsymbol{\alpha}_{jX}^{T} \mathbf{X}_{i}^{A} + \alpha_{j}G_{ij}^{A}, \label{eq:pleio1} \\
\mbox{logit}(\mu_{i'}^{B}) = \beta_{j0} + \boldsymbol{\beta}_{jX}^{T} \mathbf{X}_{i'}^{B} + \beta_{j}G_{i'j}^{B}, \label{eq:pleio2} 
\end{align}
with $\mu_{ij}^{A} = E(Y_{i}^{A} | \mathbf{X}_{i}^{A}, G_{ij}^{A})$ and $\mu_{i'j}^{B} = E(Y_{i'}^{B} | \mathbf{X}_{i'}^{B}, G_{i'j}^{B})$.
Let $Z_{j1}$ be the result of a score test for $H_{0,j,1}:\alpha_{j} = 0$ and $Z_{j2}$ the result of a score test for $H_{0,j,2}:\beta_{j} = 0$.
SNP $j$ is a pleiotropic variant associated with both lung cancer and CAD if we reject the composite null $H^{pleio}_{0,j}: \alpha_{j} = 0 \cup \beta_{j}=0$.
In massive genetic compendiums such as the UKB, A and B may also be the same dataset.
In such cases, $Z_{j1}$ and $Z_{j2}$ would generally be correlated.

A replication study is also performed with Equations (\ref{eq:pleio1}) and (\ref{eq:pleio2}) except the outcomes
$Y_{i}^{A}$ and $Y_{i'}^{B}$ are lung cancer in both models.
However, in a replication study we only want to reject the null when the parameters possess the same sign, which creates
an even larger null space than the mediation or pleiotropy composite null.
We can formalize the replication null as $H^{rep}_{0,j}: \alpha_{j} = 0 \cup \beta_{j}=0 \cup \text{sign}( \alpha_{j}) \neq \text{sign}( \beta_{j})$.
Additionally, replication and pleiotropy studies may be conducted in more than two datasets.
For instance, we may possess a third lung cancer cohort C with outcomes $Y_{i''}^{C}$ for $i''=1,...,n_{C}$ subjects 
and data on the same SNPs $\mathbf{G}_{i}^{C} = (G_{i''1}^{C},...,G_{i''J}^{C})^{T}$ 
and other covariates $\mathbf{X}_{i''}^{C} = (X_{i''1}^{C},...,X_{i''p}^{C})^{T}$. 
We can then fit $\mbox{logit}(\mu_{i''}^{C}) = \gamma_{j0} + \boldsymbol{\gamma}_{jX}^{T} \mathbf{X}_{i''}^{C} + \gamma_{j}G_{i''j}^{C}$
and let $Z_{j3}$ be the result of a score test for $H_{0,j,3}:\gamma_{j} = 0$.

For each of the mediation, pleiotropy, and replication settings, computing all the test statistics results in $J$ sets of the form
$\mathbf{Z}_{j}=(Z_{j1},...,Z_{jK})^{T}$, where each element
$Z_{jk} \sim N(0,1)$ if the $k$th null in the $j$th set is true, $k=1,...,K$.
For example, in mediation analysis $K=2$, and in three-way pleiotropy, $K=3$.
We refer to $K$ as the dimension of the composite null set. 
Let the observed value of  $\mathbf{Z}_{j}$ be given by $\mathbf{z}_{j}=(z_{j1},...,z_{jK})^{T}$.
These $\mathbf{z}_{j}$ are also called summary statistics and are often publicly available, e.g. in the GWAS catalog.

\section{Base csmGmm and Extensions}
 \label{sec:csmgmm}
 
\subsection{Multidimensional Two-Group Model}
 \label{ss:twogroup}
 
We briefly review the multidimensional two-group model to help introduce the csmGmm.
Suppose we are interested in performing large-scale testing on $J$ sets of $K$ parameters
 $\boldsymbol{\theta}_{j} = (\theta_{j1},....,\theta_{jK})^{T}$, $j=1,...,J$.
Let the general composite null for each set be $H_{0,j}^{comp,K}: \bigcup_{k=1}^{K} \theta_{jk}=0$,
and suppose the available data consists of $K$-length test statistic vectors $\mathbf{z}_{1},...,\mathbf{z}_{J}$
that have been computed as described above or provided as summary statistics.
We assume that the distribution
of $\mathbf{Z}_{j}$ depends on the unobserved true association configuration $\mathbf{H}_{j}=(H_{j1},...,H_{jK})^{T}$
\citep{efron2008microarrays, heller2014replicability}. 
Here $H_{jk}\in\{-1,0,1\}\forall j,k$, where $H_{jk}=0$ denotes
that the $k$th individual null in the $j$th set is true, i.e. $\theta_{jk}=0$.
Let $H_{jk}=-1$ denote that the $k$th individual parameter in the $j$th set is negative, i.e. $\theta_{jk}<0$,
and let $H_{jk}=1$ denote that the $k$th individual parameter in the $j$th set is positive, i.e. $\theta_{jk}>0$.
While there are technically three groups, we use the two-group name to follow historical precedence.

Let $\mathbf{h}^{l}=(h^{l}_{1},...,h^{l}_{K})^{T}\in\mathcal{H}$ denote possible values of $\mathbf{H}_{j}$, where $l=0,...,L=3^{K}-1$,
and each $l$ uniquely identifies one possible association configuration.
We use $\mathbf{h}^{0} = (0,...,0)^{T}$ to denote the global null configuration always.
The general composite null space includes all configurations with at least one 0, 
$\mathcal{H}_{0}=\{ \mathbf{h}^{l} \in \mathcal{H}: \sum_{k=1}^{K} |h^{l}_{k}| < K \}$ (replication is a special case that will be discussed later).
The local false discovery rate values are then given by
$lfdr(\mathbf{z}_{j})  :=  \text{Pr}(\mathbf{H}_{j}\in\mathcal{H}_{0} | \mathbf{Z}_{j} = \mathbf{z}_{j})$ \citep{efron2008microarrays},
which is simply 
\begin{align}
\text{Pr}(\mathbf{H}_{j}\in\mathcal{H}_{0} | \mathbf{Z}_{j} = \mathbf{z}_{j}) & = \frac{\sum_{l:\mathbf{h}^{l}\in\mathcal{H}_{0}} f(\mathbf{z}_{j} | \mathbf{H}_{j} = \mathbf{h}^{l}) \text{Pr}(\mathbf{H}_{j} =\mathbf{h}^{l}) }{\sum_{l=0}^{L}  f(\mathbf{z}_{j} | \mathbf{H}_{j} = \mathbf{h}^{l}) \text{Pr}(\mathbf{H}_{j} =\mathbf{h}^{l})}. \label{eq:lfdr}
\end{align}

It is clear that the performance of the lfdr approach depends heavily on the specifications of the densities $f(\mathbf{z}_{j} | \mathbf{H}_{j} = \mathbf{h}^{l})$.
Many options are available \citep{efron2008microarrays, heller2014replicability, wang2020imix} but may not 
(a) be flexible enough to unify the mediation, pleiotropy, and replication settings, (b) possess good operating characteristics in each application, 
or (c) provide interpretable results that do not contradict conventional
frequentist significance rankings, which is especially important in the z-statistic-dominated field of genetics research.
The csmGmm addresses these issues by building upon a Gaussian mixture model, which is known \citep{chen2009hypothesis}  for its ability to approximate
densities arbitrarily well, helping to address (a).
We then modify the mixture with conditional symmetry constraints
that greatly decrease the number of parameters, helping to address (b), and allow us to prove that (c) holds.
More specific comparisons with other density models are provided in the Supplementary Materials. 
 
\subsection{Base csmGmm}
 \label{ss:origcsmgmm}
 
 We will first consider the mediation framework with $K=2$ as a representative example to motivate and propose the csmGmm.
 The general setting of $K \geq 2$ follows.
 
 \subsubsection{Base csmGmm for $K=2$}
 \label{sss:base2}
In the classical mediation setting, there are a total of $3^2 = 9$ possible association configurations, and
five of them fall in the composite null space $\mathcal{H}_{0}^{med}=\{(0,0)^{T},(0,-1)^{T},(0,1)^{T},(-1,0)^{T},(1,0)^{T}\}$.
We index the null configurations as $l=0,1,2,3,4$, respectively.
Intuitively, the distributions of test statistics arising from $\mathbf{h}^{1}$ and $\mathbf{h}^{2}$ should be similar since
effect directions in GWAS are determined by choice of effect allele, which can vary from study to study, and 
thus the effect direction of a variant often depends on individual data cleaning pipeline choices \citep{willer2013discovery}.
The same is true of test statistics arising from $\mathbf{h}^{3}$ and $\mathbf{h}^{4}$.
This observation motivates us to define binary representations of association status as
 $b_{l}=\sum_{k=1}^{K}2^{K - k}|h^{l}_{k}|$, where $b_{l}$ is the result when the absolute values of $\mathbf{h}^{l}$ elements
 are viewed as a number in binary notation.
 
 For mediation testing, the values of $b_{l}$ match the four case numbers.
 Case 0 is the global null $\mathbf{h}^{0}=(0,0)^{T}$ with  $b_{0}=0$,
 case 1 corresponds to $\{ \mathbf{h}^{1}=(0,-1)^{T}, \mathbf{h}^{2}=(0,1)^{T} \}$ with $b_{1}=b_{2}=1$,
 case 2 corresponds to $\{ \mathbf{h}^{3}=(-1, 0)^{T}, \mathbf{h}^{4}=(1, 0)^{T} \}$ with $b_{3}=b_{4}=2$,
 and case 3 corresponds to $b_{5}=...=b_{8}=3$ for the alternative configurations $\mathcal{H}_{a}^{med}=\{(-1, -1)^{T},(-1, 1)^{T},(1, -1)^{T},(-1, 1)^{T}\}$.
To emphasize that configurations $\mathbf{h}^{l}$ with the same value of $b_{l}$ behave similarly, we describe the mixture model for $\mathbf{Z}_{j}$ as $\mathbf{Z}_{j} = \mathbf{P}_{0j} + \mathbf{P}_{1j} + \mathbf{P}_{2j} + \mathbf{P}_{3j}$; each piece corresponds to a different value of $b_{l}$.

The key characteristic of the csmGmm is to approximate the non-null dimensions of each piece by a symmetric mixture of $2M_{b_{l}}$
Gaussian distributions. 
The value of $M_{b_{l}}$ is user-determined and can be made larger for more flexibility.
Consider first $\mathbf{P}_{1j}$, which describes the case 1 configurations $\mathbf{h}^{1}$ and $\mathbf{h}^{2}$.
The mixture components belonging to this case are
\[
\mathbf{P}_{1j}  := 1\{\mathbf{H}_{j}=(0,-1)^{T}\} \sum_{m=1}^{M_{1}}1(D_{j}=m)\mathbf{Y}_{j1m} + 1\{\mathbf{H}_{j}=(0,1)^{T}\}  \sum_{m=1}^{M_{1}}1(D_{j}=m)\mathbf{Y}_{j2m},
\]
where the $\mathbf{Y}_{jlm}$ are bivariate normal random variables uniquely indexed by $(l,m)$, and
\begin{align*}
\mathbf{Y}_{jlm} & \sim MVN\{(0, h^{l}_{2} \cdot \mu_{1,m,2})^{T},\mathbf{I}_{2}\}, l=1,2, \\
\pi_{1m} & := \text{Pr} \{ \mathbf{H}_{j}=(0,-1)^{T}, D_{j} = m \}  = \text{Pr} \{ \mathbf{H}_{j}=(0,1)^{T}, D_{j} = m \}.
\end{align*}
Here $1\{\cdot \}$ is the indicator function and $\mathbf{I}_{2}$ is the $2 \times 2$ identity matrix.
We see that when the true association configuration is $\mathbf{h}^{1}$ or $\mathbf{h}^{2}$, the test statistics may come from
one of $M_{1}$ possible bivariate Gaussian distributions, which differ only in the means.
The possible mean magnitudes under $\mathbf{h}^{1}$ and $\mathbf{h}^{2}$ are the same,
and $1(D_{j}=m)$ is the indicator for the $m$th mean magnitude vector $\boldsymbol{\mu}_{1,m} = (0, \mu_{1,m,2})^{T}$, while $\mathbf{h}^{1}$ and $\mathbf{h}^{2}$ control the mean direction. 
The notation $\mu_{1,m,2}$ signifies the value in dimension $k=2$ for the $m$th possible mean magnitude vector 
when $b_{l}=1$, and the $\pi_{1m}$ are the mixing proportions for the $2M_{1}$ possible components.

In the first dimension, $\mathbf{P}_{1j}$ is equivalent to the assumption that conditional on case 1, $Z_{j1}$ is generated from
the mixture $\sum_{m=1}^{M_{1}}\pi_{1m}^{*}N(0,1)+\sum_{m=1}^{M_{1}}\pi_{1m}^{*}N(0,1)$ and thus
follows a standard $N(0,1)$ distribution.
Here $\pi_{1m}^{*}= \pi_{1m} / (2\sum_{m=1}^{M_{1}}\pi_{1m})$.
This is an obvious assumption because there is no association in $Z_{j1}$ when 
the data is generated under $\mathbf{h}^{1}$ and $\mathbf{h}^{2}$.
The second dimension of $\mathbf{P}_{1j}$ is equivalent to assuming that conditional on case 1, 
$Z_{j2}$ is generated from $\sum_{m=1}^{M_{1}}\pi_{1m}^{*}N(\mu_{1,m,2},1) + \pi_{1m}^{*}N(-\mu_{1,m,2},1)$.
Without loss of generality, we use the convention that all unknown mean parameters are non-negative.
The second dimension is symmetrical because the positive and negative directions for each mean $\mu_{1,m,2}$
have the same mixing proportion $\pi_{1m}$. 
Thus the distributions of both $Z_{j1}$ and $Z_{j2}$ are symmetric around 0 conditional on $b_{l}=1$.
The total probability of case 1 is $2\sum_{m=1}^{M_{1}}\pi_{1m}$.

The symmetry assumption is natural in many genetic association settings, and its shape is often seen in the output of 
other one-dimensional density models, even if an explicit symmetry assumption
is not made \citep{stephens2017false}.
Specifically, other density models often also assume a standard Gaussian-shaped distribution for test statistics
generated under the null, with a symmetric bimodal distribution for test statistics generated under the alternative \citep{stephens2017false}.
In the csmGmm, the conditional symmetry assumption is critical for reducing the number of parameters,
thus enabling (b), and for enforcing (c).

The piece $\mathbf{P}_{2j}$ follows naturally from the form of $\mathbf{P}_{1j}$.
The case 2 configurations $\mathbf{h}^{3}=(-1,0)^{T}$ and $\mathbf{h}^{4}=(1,0)^{T}$
are similar to those with $b_{l}=1$, except the roles of the first and second dimension are switched.
We have $\mathbf{P}_{2j} := 1\{\mathbf{H}_{j}=(-1, 0)^{T}\} \sum_{m=1}^{M_{2}}1(D_{j}=m)\mathbf{Y}_{j3m} + 1\{\mathbf{H}_{j}=(1, 0)^{T}\}  \sum_{m=1}^{M_{2}}1(D_{j}=m)\mathbf{Y}_{j4m}$, where $\mathbf{Y}_{jlm} \sim MVN\{ (h^{l}_{1} \cdot \mu_{2,m,1}, 0)^{T},\mathbf{I}_{2}\}, l=3,4$ and 
$\pi_{2m}  := \text{Pr} \{ \mathbf{H}_{j}=(-1,0)^{T}, D_{j} = m \}  = \text{Pr} \{ \mathbf{H}_{j}=(1,0)^{T}, D_{j} = m \}.$
Thus conditional on case 2, the second dimension is $N(0,1)$, and the first dimension is generated from
the symmetric mixture $\sum_{m=1}^{M_{2}}\pi_{2m}^{*}  N(\mu_{2,m,1},1) + \pi_{2m}^{*}  N(-\mu_{2,m,1},1) $, with
$\pi_{2m}^{*} = \pi_{2m} / (2\sum_{m=1}^{M_{2}}\pi_{2m})$.
Here the $\mu_{2,m,1}$ denotes the value in dimension $k=1$ for the $m$th possible mean magnitude vector 
when $b_{l}=2$, and $\pi_{2m}$ are the mixing proportions for $b_{l}=2$.

The piece $\mathbf{P}_{0j}$ corresponds to the global null configuration $\mathbf{h}^{0}$ with $b_{0}=0$ (case 0) and is simply the standard bivariate Gaussian
\begin{align*}
\mathbf{P}_{0j}  := & 1\{\mathbf{H}_{j}=(0,0)^{T}\}\mathbf{Y}_{j01},\\
\mathbf{Y}_{j01}  \sim & MVN\{(0,0)^{T},\mathbf{I}_{2}\}, \\
\pi_{01} & := \text{Pr} \{ \mathbf{H}_{j}=(0,0)^{T} \}, 
\end{align*}
as would be expected when there is no association in either dimension.
The global null portion trivially has $M_{0}=1$ always.
Hence whenever $\mathbf{Y}_{j0m}$ and $\pi_{0m}$ appear, they are always 
referring to $\mathbf{Y}_{j01}$ and $\pi_{01}$, with the latter being
the probability of case 0.

Finally, we consider the composite null alternative portion $\mathbf{P}_{3j}$. 
This piece models the configurations with $b_{l}=3$ and follows the same
conditionally symmetric structure:
\begin{align*}
\mathbf{P}_{3j} & :=  1\{\mathbf{H}_{j}=(-1,-1)^{T}\}  \sum_{m=1}^{M_{3}}1(D_{j}=m)\mathbf{Y}_{j5m} + 1\{\mathbf{H}_{j}=(-1,1)^{T}\}  \sum_{m=1}^{M_{3}}1(D_{j}=m)\mathbf{Y}_{j6m} +  \\
& 1\{\mathbf{H}_{j}=(1,-1)^{T}\}  \sum_{m=1}^{M_{3}}1(D_{j}=m)\mathbf{Y}_{j7m} + 1\{\mathbf{H}_{j}=(1,1)^{T}\}   \sum_{m=1}^{M_{3}}1(D_{j}=m)\mathbf{Y}_{j8m}.
\end{align*}
The $4M_{3}$ terms described above reduce to only $2M_{3}$ distinct components in each dimension because of familiar constraints: 
\begin{align}
\mathbf{Y}_{jlm} & \sim MVN\{(h^{l}_{1} \cdot \mu_{3,m,1}, h^{l}_{2} \cdot \mu_{3,m,2})^{T},\mathbf{I}_{2}\},  l=5,6,7,8, \nonumber \\
\pi_{3m} & := \text{Pr} \{ \mathbf{H}_{j}=(-1,-1)^{T}, D_{j} = m \}  =  \text{Pr} \{ \mathbf{H}_{j}=(-1,1)^{T}, D_{j} = m \} \nonumber \\
& = \text{Pr} \{ \mathbf{H}_{j}=(1,-1)^{T}, D_{j} = m \} = \text{Pr} \{ \mathbf{H}_{j}=(1,1)^{T}, D_{j} = m \}, \nonumber \\
\mu_{3,m,k} & \geq \mu_{b,m',k}>0  \text{ } \forall b<3 \text{ and } \forall m,m',k.  \label{eq:muassumption}
\end{align}
As above, we can see that conditional on $b_{l}=3$,  $Z_{j1}$ is generated from the symmetric mixture 
$\sum_{m=1}^{M_{3}}2\pi_{3m}^{*} N(-\mu_{3,m,1},1)+2\pi_{3m}^{*}N(\mu_{3,m,1},1)$, 
and the $Z_{j2}$  is similarly generated from the symmetric mixture
$\sum_{m=1}^{M_{3}}2\pi_{3m}^{*} N(-\mu_{3,m,2},1)+2\pi_{3m}^{*}N(\mu_{3,m,2},1)$, where $\pi_{3m}^{*} = \pi_{3m} / (4\sum_{m=1}^{M_{3}}\pi_{3m})$.
The assumption (\ref{eq:muassumption}) is necessary for (c) and means that for any dimension $k$, the possible mean values
in that dimension for configurations in $\mathcal{H}_{a}^{med}$ must not be smaller than any of the possible mean values for configurations in $\mathcal{H}_{0}^{med}$.
For example, if $M_{1}=M_{2}=M_{3}=1$, it must be true that $\mu_{3,1,1} \geq \mu_{2,1,1}$ and $\mu_{3,1,2} \geq \mu_{1,1,2}$.
This is a plausible assumption in many cases, for instance a
SNP that is associated with many different diseases may encode critical
functions with large effect sizes compared to SNPs that perturb risk
only in one disease.

 \subsubsection{Base csmGmm in general $K\geq2$ settings}
 \label{sss:basegeneral}
 
For general composite null settings, which may involve $K>2$ total dimensions, the model
for $\mathbf{Z}_{j}$ conditional on an association configuration $\mathbf{h}^{l}$ can be compactly
represented without describing each $b_{l}$ piece separately.
The conditional distribution is a mixture of $M_{b_{l}}$ multivariate Gaussian random variables
\[
\mathbf{Z}_{j}|\mathbf{H}_{j}=\mathbf{h}^{l}\overset{d}{=}\sum_{m=1}^{M_{b_{l}}}1(D_{j}=m)\mathbf{Y}_{jlm},
\]
where $l=0,...,3^{K}-1$, $\overset{d}{=}$ stands for equivalence in distribution, and
\begin{align}
\label{eq:csmGmm}
\begin{split}
\mathbf{Y}_{jlm} & \sim MVN \{ \text{diag}(\mathbf{h}^{l}) \boldsymbol{\mu}_{b_{l},m},\mathbf{I}_{K} \},   \\
\boldsymbol{\mu}_{b,m} & =(\mu_{b,m,1},...,\mu_{b,m,K})^{T}, b=0,1,...,2^{K}-1, \\
\mu_{2^{K}-1,m,k} & \geq \mu_{b,m',k}>0  \text{ } \forall b<2^{K}-1 \text{ and } \forall m,m',k,  \\
\pi_{b_{l}m} & := \text{Pr} ( \mathbf{H}_{j}=\mathbf{h}^{l}, D_{j} = m ) = \text{Pr} ( \mathbf{H}_{j}=\mathbf{h}^{l'}, D_{j} = m ) \text{ if } b_{l} = b_{l'}.
\end{split}
\end{align}

A few notes may help clarify some of the busy notation. 
The joint probabilities $\pi_{b_{l}m} =\text{Pr} ( \mathbf{H}_{j}=\mathbf{h}^{l}, D_{j} = m )$ have the property that
$\sum_{l=0}^{3^{K}-1} \sum_{m=1}^{M_{b_{l}}} \pi_{b_{l}m} = 1$, as the mixing proportions for each component must sum
to unity.
Also, terms $\mu_{b,m,k}$ refer to the mean magnitude in dimension $k$ for the $m$th possible magnitude vector $\boldsymbol{\mu}_{b,m}$ 
used by configurations with $b_{l}=b$.
Finally, the compacted notation above does not explicitly show means of 0, but it is understood that $\mu_{b,m,k}=0$  when configurations with binary representation $b$ have a 0 in the $k$th dimension. 
The 0s can be easily determined by writing out $b$ in binary notation.
For example, in mediation (Section \ref{sss:base2}), when $b=1$ we can write this
in binary notation as 01, showing that there is no association in the first dimension $k=1$. 
Thus $\mu_{1,m,1}=0\ \forall m$.

While the terms have been compressed, the csmGmm structure remains the same.
The vector $\mathbf{Z}_{j}$ is still a mixture of multivariate normal Gaussians components $Y_{jlm}$.
These $Y_{jlm}$ components are again uniquely determined by the values of $(\mathbf{H}_{j},D_{j})$,
which describe both the association configuration and the magnitude of the $Y_{jlm}$ mean.
Most importantly, the densities in each dimension are still symmetric conditional on the value of $b_{l}$.
For example, in three-dimensional applications with $\mathbf{Z}_{j}=(Z_{j1}, Z_{j2}, Z_{j3})^{T}$, there
are 8 alternative configurations with $b_{l}=7$.
Conditional on $b_{l}=7$, the density of $\mathbf{Z}_{j}$ looks like $\mathbf{P}_{3j}$, except each term in $\mathbf{P}_{3j}$
splits symmetrically into two terms, one to account for a negative mean in dimension $k=3$ and one to account for a positive mean.
The conditional symmetry continues to scale similarly in higher dimensions because the mean magnitude vectors and mixing proportions
are shared among all configurations within a binary representation.

Let $\boldsymbol{\mu}=(\boldsymbol{\mu}_{0}, \boldsymbol{\mu}_{1},...,\boldsymbol{\mu}_{2^{K}-1})$
be the collection of all means, where $\boldsymbol{\mu}_{b}=(\boldsymbol{\mu}_{b,1},...,\boldsymbol{\mu}_{b,M_{b}}).$ 
Additionally, let the full vector of probabilities be held in $\boldsymbol{\pi}=(\pi_{0},\boldsymbol{\pi}_{1},...,\boldsymbol{\pi}_{2^{K}-1})^{T}$
where $\boldsymbol{\pi}_{b} = (\pi_{b1},...,\pi_{bM_{b}})^{T}$.
When $\boldsymbol{\pi}$ and $\boldsymbol{\mu}$ are known, the local false discovery rate under the base csmGmm in (\ref{eq:csmGmm}) can then be calculated as
\begin{align}
   \text{Pr}(\mathbf{H}_{j}\in\mathcal{H}_{0}|\mathbf{z}_{j}) =  \frac{\sum_{l:\mathbf{h}^{l}\in\mathcal{H}_{0}}\sum_{m=1}^{M_{b_l}}\pi_{b_{l}m}\prod_{k=1}^{K}\phi(z_{jk} - h^{l}_{k} \mu_{b_{l},m,k})}{\sum_{l=0}^{L}\sum_{m=1}^{M_{b_l}}\pi_{b_{l}m}\prod_{k=1}^{K}\phi(z_{jk} - h^{l}_{k} \mu_{b_{l},m,k})}. \label{eq:lfdr_csmGmm} 
\end{align}
Here $\phi(\cdot)$ is the density of a standard Gaussian random variable.

Crucially, we can show that the csmGmm guarantees the lfdr-value testing rule will not contradict conventional 
 z-statistic testing rules.
To be more precise, we define the concept of an incongruous Bayesian result when 
performing composite null inference under the two-group model.

\begin{definition} 
Suppose we are performing large-scale composite null inference
using the previously-defined data $\mathbf{z}_{1},...,\mathbf{z}_{J}$.
Assume that inference is performed by applying the two-group model to each $\mathbf{z}_{j}$ to calculate lfdr-values as in Equation (\ref{eq:lfdr}).
This analysis produces an incongruous result for $\mathbf{z}_{j}$ 
if the lfdr-value for $\mathbf{z}_{j}$ is greater than the lfdr-value for another set $\mathbf{z}_{j'}$
when $\text{sign}(z_{jk}) = \text{sign}(z_{j'k}) \forall k$ and $|z_{jk}| \geq |z_{j'k}| \forall k$.
\end{definition} 

Theorem 1 shows that the csmGmm prevents incongruous results. 
\begin{theorem}  Suppose we are performing composite null inference on $H_{0,1}^{comp,K},..., H_{0,J}^{comp,K}$
using the data $\mathbf{z}_{1},...,\mathbf{z}_{J}$.
When utilizing the two-group model with the csmGmm described in Equation (\ref{eq:csmGmm}),
for any given $k$, the lfdr-value of Equation (\ref{eq:lfdr}) is non-increasing in $z_{jk}$ when $z_{jk}$ is positive and
the other elements are held fixed. 
Additionally,  the lfdr-value is non-decreasing in $z_{jk}$
when $z_{jk}$ is negative and the other elements are held fixed.
Thus, the csmGmm never produces incongruous results.
This property holds for sets of any size $K \geq 2$.
\end{theorem}

We see that the csmGmm prevents contradictions between the empirical Bayes and frequentist significance rankings
in mediation analysis.
The same is true for pleiotropy analysis on independent datasets, even when there are more than two dimensions, $K>2$.
The proof of Theorem 1 is provided the Supplementary Materials.

\subsection{Correlation Within Sets}
 \label{ss:ccsmgmm}

In a pleiotropy analysis conducted within a single dataset, e.g. the UK Biobank, test statistics in a set may be correlated. 
As the base csmGmm assumes independence between test statistics in the same set, it would
be a poor fit for such correlated situations.
Thus, we propose a correlated csmGmm (c-csmGmm) model for sets of dimension $K=2$.
Suppose the correlation between the test statistics is known or can be well-estimated
as $\rho$.
For instance, for two genome-wide association studies, one can simply use the 
sample correlation of the test statistics \citep{liu2018multiple} to estimate $\rho$.
Then, conditional on the hypothesis configuration $\mathbf{h}^{l}$, the model for $\mathbf{Z}_{j}$ is 
\begin{align}
\label{eq:correlatedModel}
\begin{split}
 \mathbf{Z}_{j}|\mathbf{H}_{j} =\mathbf{h}^{l} & \overset{d}{=}Y_{jl},\\
\mathbf{Y}_{jl} & \sim MVN\left(\text{diag}\{\mathbf{h}^{l}\}\boldsymbol{\mu}_{b_{l}},\left(\begin{array}{cc}
1 & \rho\\
\rho & 1
\end{array}\right)\right), \\
\boldsymbol{\mu}_{b_{l}} & =(\mu_{b_{l},1},\mu_{b_{l},2})^{T},\\
\mu_{3,k} & \geq \mu_{b,k}>0 \text{ } \forall b<3 \text{ and } k=1,2.
\end{split}
\end{align}
Model (\ref{eq:correlatedModel}) assumes two-dimensional composite nulls and $M_{1}=M_{2}=M_{3}=1$.
Here, similar to Section \ref{sss:basegeneral}, we let $\mu_{b,k}=0$  when configurations with binary representation $b$ have a 0 in the $k$th dimension. 
We can then prove a result about incongruous results similar to Theorem 1 for the c-csmGmm as well.

\begin{theorem} 
Suppose we are performing composite null inference on $H_{0,1}^{comp,2},..., H_{0,J}^{comp,2}$
using the bivariate vectors $\mathbf{z}_{1},...,\mathbf{z}_{J}$.
When utilizing the two-group model with the c-csmGmm proposed in Equation (\ref{eq:correlatedModel}) with $\rho\geq0$,
for either $k$, the lfdr-value of Equation (\ref{eq:lfdr}) is non-increasing in
$z_{jk}$ when the other element of $\mathbf{z}_{j}$ is held fixed and $z_{jk} > c_{p}$ for some 
positive constant $c_{p}$.
Additionally, the lfdr-value is non-decreasing in $z_{jk}$ when the other element of $\mathbf{z}_{j}$ is held fixed
and $z_{jk}<c_{n}$ for some negative constant $c_{n}$. 
The constants $c_{p}$ and $c_{n}$ depend on the correlation structure and are described in the Supplementary Materials.
\end{theorem}

In other words, the c-csmGmm prevents incongruous results in the tail. 
This behavior is desirable since the tail is generally the area of interest, where we would expect to find significant discoveries.
Incongruous results for test statistics near 0 may possess little impact on the final conclusions of an analysis.
The condition that $\rho \geq 0$ is also not restrictive as generally pleiotropy studies are conducted on 
diseases that are positively correlated with each other.
We prove Theorem 2 in the Supplementary Materials.
The approach is substantially more complicated than Theorem 1, as the correlation adds complexity to
each joint density.

\subsection{Replication Analysis}
 \label{ss:rcsmgmm}
 
Replication analyses also require slightly different considerations because of the different form of the composite null.
In a replication analysis, if we are again testing the parameters $\boldsymbol{\theta}_{j}$ with $\mathbf{z}_{j}$,
the general replication null is $H_{0, j}^{rep, K}: \{ \bigcup_{k=1}^{K} \theta_{jk} = 0 \} \cup \{ \bigcup_{k\neq k'} \mbox{sign} (\theta_{jk}) \neq \mbox{sign} (\theta_{jk'})  \}$.
Correspondingly, we need to modify the null and alternative spaces in the two-group model,
with $\mathcal{H}^{rep}_{a}=\{(1,1,...,1)^{T}_{K\times 1},(-1,-1,...,-1)^{T}_{K\times 1}\}$ and $\mathcal{H}^{rep}_{0} = \mathcal{H} \backslash \mathcal{H}^{rep}_{a}$.

The elements of  $\mathbf{z}_{j}$ will still be mutually independent, and 
only a slight modification of the base csmGmm is needed to construct the replication csmGmm (r-csmGmm).
Specifically, the r-csmGmm is defined as model  (\ref{eq:csmGmm}) with the additional restriction that $M_{2^K-1} = 1$.
In other words, conditional on alternative configurations with $b_{l}=2^{K}-1$, the density of each
dimension must be a mixture of two Gaussian distributions with the same mean magnitude and opposite directions (see example of Section \ref{sss:base2}).
This constraint is reasonable when test statistics under the alternative could plausibly exhibit
such a bimodal symmetric shape.
The assumption may not be appropriate if test statistics under the alternative are believed to possess a density
with more than two peaks.
We then have the following theorem.

\begin{theorem}  
Suppose we are performing replication composite null inference on $H_{0,1}^{rep,K},..., H_{0,J}^{rep,K}$
using the data $\mathbf{z}_{1},...,\mathbf{z}_{J}$ and the replication null space $\mathcal{H}^{rep}_{0}$.
When utilizing the two-group model with the r-csmGmm,
for any given $k$, the lfdr-value of  Equation (\ref{eq:lfdr}) is non-increasing in $z_{jk}$ when $z_{jk}$ is positive and
the other elements are held fixed as long as $z_{jk'} > 0 \forall k'$. 
Additionally, the lfdr-value is non-decreasing in $z_{jk}$ when
$z_{jk}$ is negative and the other elements are held fixed as long
as $z_{jk'} < 0 \forall k'$. 
This property holds for replication composite null sets of any size $K \geq 2$.
\end{theorem}

Theorem 3 differs from Theorem 1 in that congruous results are only guaranteed when all observed test statistics possess the same sign. 
This condition is not restrictive as observed test statistics with different signs generally would not be considered significant in a replication analysis anyway.
We prove Theorem 3 in the Supplementary Materials.


\section{Empirical Bayes Analysis}
\label{sec:empbayes}
 
 \subsection{Parameter Estimation}
\label{ss:estimation}
 The full data likelihood of $\mathbf{z}_{1},...,\mathbf{z}_{J}$ is very complex in genetic
 association studies due to correlation between SNPs.
 We can instead consider the composite likelihood ignoring correlations among features $j'\neq j$,
 which has been justified in local dependency settings \citep{lindsay1988composite, heller2014replicability}.
This composite likelihood is 
\begin{align}
\mathcal{L} & =\prod_{j=1}^{J}\prod_{\mathbf{c}^{l,m}\in\mathcal{C}}\left\{  \pi_{b_{l}m} f(\mathbf{Z}_{j}| \mathbf{C}_{j} = \mathbf{c}^{l,m} )\right\} ^{1(\mathbf{C}_{j}=\mathbf{c}^{l,m})},
\label{eq:likelihood}
\end{align}
where $\mathbf{C}_{j}=(\mathbf{H}_{j}^{T},D_{j})^{T}$ identifies the assocation configuration and mean vector,
$\mathbf{C}_{j}=\mathbf{c}^{l,m}$ denotes that $\mathbf{C}_{j}=((\mathbf{h}^{l})^{T},m)^{T}$,
and $\mathcal{C}$ is the collection of all possible mixture components.
The log-likelihood is then
\begin{small}
\[
\ell = \sum_{j=1}^{J}\sum_{\mathbf{c}^{l,m}\in\mathcal{C}}1(\mathbf{C}_{j}=\mathbf{c}^{l,m})\left[\log\left( \pi_{b_{l}m} \right)-\frac{1}{2}(\mathbf{z}_{j}-\boldsymbol{\Lambda}_{l}\boldsymbol{\mu}_{b_{l},m})^{T}\boldsymbol{\Sigma}^{-1}(\mathbf{z}_{j}-\boldsymbol{\Lambda}_{l}\boldsymbol{\mu}_{b_{l},m})\right],
\]
\end{small}
where $\boldsymbol{\Lambda}_{l} := \text{diag} \{\mathbf{h}^{l}\}$, and $\boldsymbol{\Sigma}$ is either
the identity matrix or correlation matrix specified in the c-csmGmm, depending on the setting.
We can then estimate the unknown parameters $\boldsymbol{\pi}$ and $\boldsymbol{\mu}$
using a standard expectation-maximization (EM) algorithm \citep{dempster1977maximum}.

 \subsection{Model Fit and Inference}
\label{ss:inference}
Much literature has investigated the convergence of the EM algorithm and other technical issues in mixture distributions \citep{chen2009hypothesis}. 
The full details of our implementation are provided in the Supplementary Materials, 
as we note that the focus of this work is performing inference on the $\textbf{z}_{j}$.
We do highlight that effect directions may not appear symmetrically distributed around 0 for certain studies.
For instance, some lipids summary statistics are oriented such that all effects are positive
\citep{willer2013discovery}.
However, using all positive orientations is clearly an artificial decision, and no information is lost by simply
switching effect and non-effect alleles, which negates the test statistic.
When test statistics appear skewed to one side, simulations show that randomly performing effect-reference switching to achieve 
rough symmetry about the origin works well (see Supplementary Materials).

To perform inference, we can apply the standard result that the mean lfdr-value in a rejection region 
is the probability of a false rejection \citep{efron2002empirical, heller2014replicability}.
Specifically, we define the rejection region for a nominal false discovery rate $q$ to be
 $\mathcal{R}_{q}:=\{\mathbf{z}_{j}:\widehat{lfdr}(\mathbf{z}_{j})\leq  \widehat{lfdr}_{(\hat{t}(q))}   \}$.
Here $\widehat{lfdr}_{(j)}$ denotes the order statistics of the $\widehat{lfdr}(\mathbf{z}_{j})$, 
$\widehat{lfdr}(\mathbf{z}_{j})$ are the lfdr-values computed under Equation (\ref{eq:lfdr}) and a csmGmm model
with estimates $(\hat{\boldsymbol{\mu}}, \hat{\boldsymbol{\pi}})$ in place of  $(\boldsymbol{\mu}, \boldsymbol{\pi})$, and
$\widehat{t}(q)=\max_{r}\sum_{j=1}^{r} \widehat{lfdr}_{(j)}/r\leq q$ is the largest $r$ such that an average of
the $r$-smallest lfdr-values is less than or equal to $q$.


\section{Simulation}
 \label{sec:simulation}
 
We next study the operating characteristics of various composite null hypothesis testing methods through simulation.
Comparisons are made against the divide and aggregate composite null test (DACT) \citep{liu2022large}, the 
high-dimensional mediation test (HDMT) \citep{dai2022multiple}, 
and the empirical Bayes two-group approach with three other density models:
\verb|locfdr| package natural splines with default 7 degrees of freedom  \citep{heller2014replicability}, 
natural splines with 50 degrees of freedom, and a standard Gaussian kernel density estimator.
For the HDMT we use the default fdr-estimation mode, and for the
DACT we use the default frequentist mode with the standard Benjamini-Hochberg method \citep{benjamini1995controlling} to control false discovery rate.
The nominal false discovery rate is always $q=0.1$.

All tests generally perform well in the most basic mediation-type settings when signals are sparse
and non-zero effect sizes are weak.
When the proportion of composite alternative sets is large, when large effect sizes exist in both
the composite null and composite alternative, or in higher-dimensional settings, the DACT and two-group approaches with other density models are
not always able to protect the nominal error rate, as the density shapes become more complex.
The HDMT is able to control false discovery rates in a range of settings, however it can also be too conservative \citep{dai2022multiple}
and is not applicable outside of the mediation setting.
The csmGmm not only controls false discoveries in a variety of situations but also offers reliable
power and can be utilized in a larger number of settings, e.g. to test three-dimensional
composite null hypotheses or in replication analyses.
For sake of space, we present only some main results and offer further details in the Supplementary Materials.

 \subsection{Two-Dimensional Mediation}
 \label{ss:medsims}
 
\begin{figure}[!ht]
\begin{center}
\includegraphics[scale=0.31, angle=0]{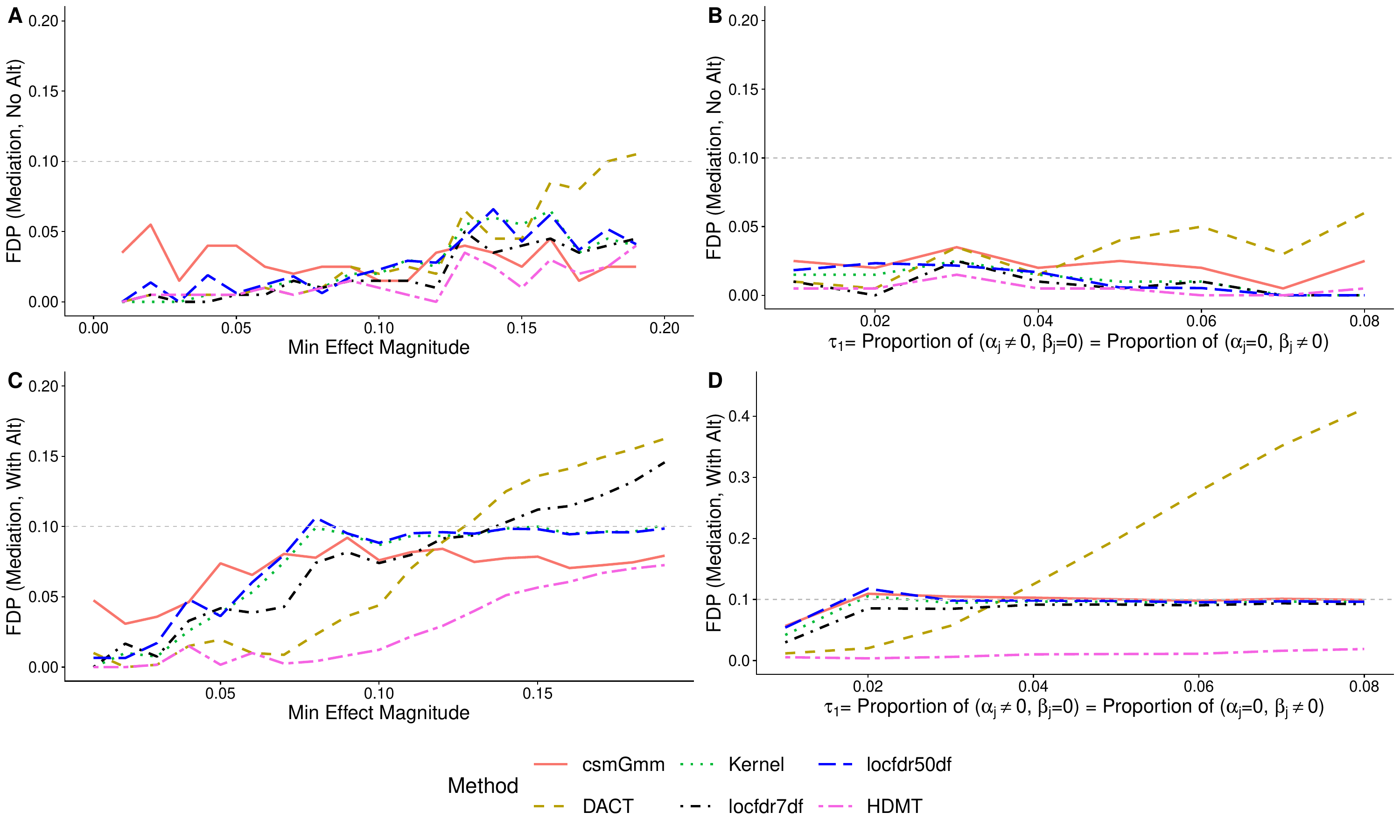}  
\end{center}
\caption{False discovery proportions for mediation settings when all data is generated under the composite null (A-B) or when some data is generated under the composite alternative (C-D).
When effect magnitudes are varied (A and C), the magnitudes of $\alpha_{j}$ for all cases are generated from a uniform distribution of length 0.1, with the lower end of the window
given on the x-axis. 
The magnitudes of $\beta_{j}$  are generated similarly and then 0.04 is added so that test statistics have means of approximately 5 and 4 at the right end of the x-axis.
Also, A and C set $\tau_{1} =0.02$. 
When $\tau_{1}$ is varied (B and D), non-zero magnitudes are $\alpha_{j} = 0.14$ and $\beta_{j} = 0.18$ always.
In panels A and B, $\tau_{2}=0$, and in panels C and D we set $\tau_{2}=3\tau_{1}^2$.
Each iteration is conducted with $10^5$ sets of test statistics, each plotted point reflects the average of 200 iterations, and points are plotted at 0.01 intervals
on the x-axis.}
\label{fig:med_fdp}
\end{figure}

Figure \ref{fig:med_fdp} considers the original csmGmm in the basic two-dimensional mediation setting.
Each simulation iteration generates $j=1,..., 10^{5}$ sets of two test statistics by simulating under
$M_{ij} = \alpha_{0} + \alpha_{j} G_{ij} + \epsilon_{ij}$ and $\mbox{logit}(\mu_{ij}) = \beta_{0} + \beta_{j} M_{ij} + \beta_{G} G_{ij}$.
Here  $\mbox{logit}(\mu_{ij}) = E(Y_{ij} | M_{ij}, G_{ij})$, $G_{ij}$ is a Binomial$(2, 0.3)$ random variable, $(\alpha_{0}, \beta_{0}, \beta_{G}) = (0, -1, 0)$, 
$\epsilon_{ij} \overset{iid}{\sim} N(0,1)$, and $n=1000$.
Standard score statistics for $(\alpha_{j}, \beta_{j})$ are calculated by fitting the same regression models.
We vary both the proportion of $(\alpha_{j}, \beta_{j})$ that are non-zero and their effect sizes; effects are split evenly
 between positive and negative directions (see Supplementary Materials for all positive case).
Figures \ref{fig:med_fdp}A and \ref{fig:med_fdp}B consider the composite null setting where
there are no true mediated effects, while Figures \ref{fig:med_fdp}C and \ref{fig:med_fdp}D
include a non-zero proportion of mediated effects (case 3), specifically $\tau_{2} = 3\tau_{1}^{2}$. 
Here $\tau_{1}$ is the common proportion of each configuration with only one
true association (cases 1 and 2).

These simulations demonstrate two major points about the mediation setting.
First, when $\tau_{2}=0$ and there are no composite null sets generated under the alternative (Figures \ref{fig:med_fdp}A and \ref{fig:med_fdp}B),
all methods can protect the false discovery rate well, and all are generally conservative.
This finding confirms previous reports \citep{dai2022multiple, liu2022large}.
Second, when some sets are generated under the alternative, all methods can protect
the nominal error rate if $\tau_{1}$ (and thus $\tau_{2}$) is small, or if associations throughout the dataset are small.
However if effects are generally large (Figure \ref{fig:med_fdp}C) or $\tau_{1}$ is large (Figure \ref{fig:med_fdp}D), some approaches may be anticonservative.
In Figure \ref{fig:med_fdp}C we see that both the 7 degrees of freedom
spline and the DACT are anticonservative when associations generate test statistics with means in the range of 3 to 4 ($x\approx0.15$).
The DACT was developed for rare-weak mediation effects, and our simulations show that it performs well in that setting.
Although the HDMT protects the nominal false discovery rate, it is much more conservative than other methods,
including when $\tau_{1}$ is large.
The csmGmm generally protects the nominal false discovery rate well even in more extreme settings and
is less conservative than the HDMT.

\begin{figure}[!ht]
\begin{center}
\includegraphics[scale=0.33, angle=0]{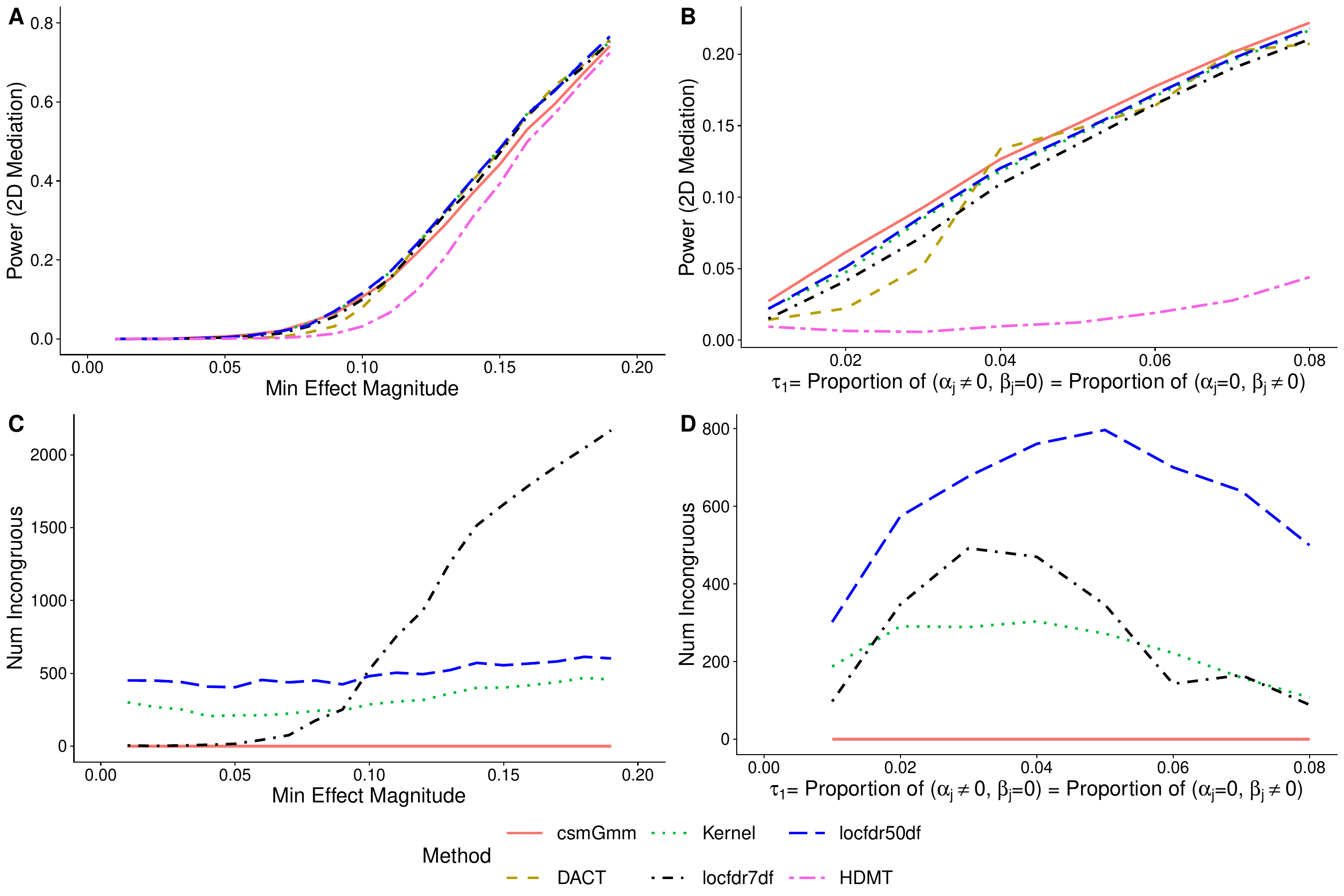}  
\end{center}
\caption{Power (A-B) and number of incongruous effects (C-D) for mediation settings.
The data in A and C is generated as in Figure \ref{fig:med_fdp}C, and the data in B and D
is generated as in Figure \ref{fig:med_fdp}D.
The csmGmm shows comparable power to other approaches (see Supplementary Materials for details
about power simulation), while the HDMT shows 
considerable power loss in some settings.
The DACT and HDMT do not calculate lfdr-values and so do not produce incongruous results.}
\label{fig:med_power}
\end{figure}

Figures \ref{fig:med_power}A and  \ref{fig:med_power}B further show test power 
under the settings of Figures \ref{fig:med_fdp}C and \ref{fig:med_fdp}D, respectively.
We can see that almost all methods possess the same power, with the csmGmm slightly
underperforming in Figure \ref{fig:med_power}A while providing the best power in Figure \ref{fig:med_power}B (see Supplementary Materials
for details about power simulations).
Only the HDMT lags behind other tests in these simulations, which is expected \citep{dai2022multiple} given the conservativeness shown in Figures \ref{fig:med_fdp}C-D.
Figures \ref{fig:med_power}C and \ref{fig:med_power}D show that, as previously proved, the csmGmm does not produce incongruous results, 
while the other empirical Bayes
approaches can produce hundreds or even thousands of incongruous results.

 \subsection{Three-Dimensional Pleiotropy}
 \label{ss:pleiosims}
 
We investigate larger composite null sets with a three-dimensional pleiotropy simulation similar to the lung cancer analysis.
For each $j$, three outcomes $(Y_{ij1}, Y_{i'j2}, Y_{i''j3})$ are generated independently 
under $\mbox{logit}(\mu_{ij1}) = \alpha_{0} + \alpha_{j} G_{ij}$, $\mbox{logit}(\mu_{i'j2}) = \beta_{0} + \beta_{j} G_{i'j}$,
and $\mbox{logit}(\mu_{i''j3}) = \gamma_{0} + \gamma_{j} G_{i''j}$.
Here, $\mu_{ij1}=E(Y_{ij1})$, $\mu_{i'j2}  = E(Y_{i'j2})$,  $\mu_{i''j3}  = E(Y_{i''j3})$, the genotypes are again Binomial$(2, 0.3)$, and 
the intercepts are $\alpha_{0} = \beta_{0} = \gamma_{0} = -1$.
We calculate standard score statistics for $(\alpha_{j}, \beta_{j}, \gamma_{j})$ by fitting the same regression models.
The HDMT and DACT are not available for three-dimensional settings and are not applied.

\begin{figure}[!ht]
\begin{center}
\includegraphics[scale=0.32, angle=0]{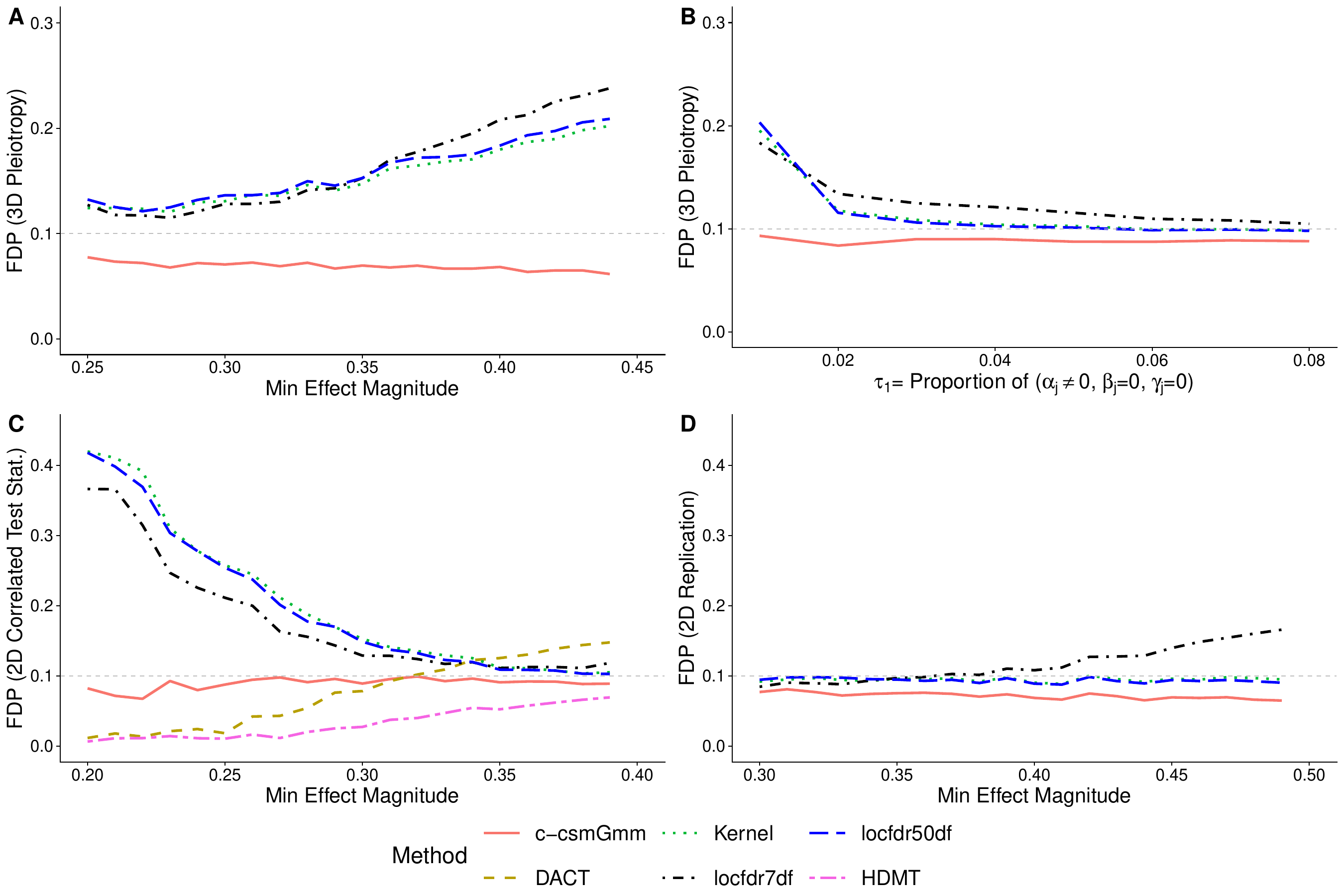}  
\end{center}
\caption{False discovery proportions for three-dimensional pleiotropy setting (A-B), correlation within sets setting (C),
and replication setting (D). When three-dimensional effect magnitudes are varied (A), we set $\tau_{3} = \tau_{2}/2, \tau_{2}=3\tau_{1}^{2},\tau_{1}=0.02$.
The left and right ends of the x-axis correspond to test statistics with absolute mean of approximately 4 and 6.5, respectively (see Supplementary Materials
for power) due to increasing difficulty of modeling three-dimensional densities.
When $\tau_{1}$ is varied (B), we set non-zero effect magnitudes at 0.38. 
In two-dimensions (C-D), the proportions of non-zero effects are set as in Figure \ref{fig:med_fdp}C.}
\label{fig:sim3d}
\end{figure}

Figures \ref{fig:sim3d}A-B demonstrate the effects of varying effect sizes and proportion of non-zero effects,
and all data includes composite alternatives generated at a proportion of $\tau_{3} = \tau_{2} / 2, \tau_{2} = 3\times \tau_{1}^2$,
where $\tau_{2}$ and $\tau_{1}$ are the common proportions of composite null configurations with two and one association, respectively.
The patterns that appear are similar to those from Figures \ref{fig:med_fdp} and \ref{fig:med_power}.
Again, the spline and kernel density approaches can be anticonservative in certain situations. 
In three dimensions, this poor performance can occur when effect sizes are large  (Figure \ref{fig:sim3d}A) or when the proportion of non-zero effects is small
(Figure \ref{fig:sim3d}B).
In the former case, composite null sets with one or two associations can appear similar to alternative sets due to 
the large magnitude of associations, and in the latter case, the rarity of composite alternative sets makes them hard to 
differentiate from null sets.
Both settings create challenges for density estimation.
More details about power and other three-dimensional results are available in the Supplementary Materials.

 \subsection{c-csmGmm and r-csmGmm}
 \label{ss:othersims}
 
A final set of simulations considers use of the c-csmGmm and r-csmGmm.
When generating correlated pleiotropy test statistics, we use the models $\text{logit}(\mu_{ij1}) = \alpha_{0} + \alpha_{j} G_{ij}$ and $\text{logit}(\mu_{ij2}) = \beta_{0} + \beta_{j} G_{ij}$,
where the $G_{ij} \sim \text{Binomial}(2,0.3)$ is the same for both outcomes, and the two outcomes $Y_{ij1}$ and $Y_{ij2}$ are correlated at $\rho = 0.1$.
This situation imitates our analysis of lung cancer and CAD within the UK Biobank.
For replication simulations, we generate data in the same way except in two separate cohorts, with uncorrelated outcomes.

Figure \ref{fig:sim3d}C shows that the c-csmGmm performs well both when effect sizes
are small and when they are large.
The HDMT is conservative, as we have observed in each other setting.
All other procedures may be anticonservative when effect sizes are too small or too large.
However it should be noted that outside of the c-csmGmm, none of the other approaches
were developed for testing sets with correlated test statistics, and it may be unfair
to expect them to perform well in this setting.
Simulations investigating correlation between SNPs, with the use of real genotypes, are available in the Supplementary Materials.

Figure \ref{fig:sim3d}D shows that the r-csmGmm is slightly conservative when testing
for replication.
In fact, we observe that the 50 degrees of freedom spline and kernel density estimate both perform better at adhering to the nominal false discovery rate.
Such a finding is not surprising as the multivariate two-group model using \verb|locfdr| was originally proposed for the replication setting \citep{heller2014replicability}
and would thus be expected to perform well in this application.
Differences in relative performance between the replication setting and other settings may be due to the different number of configurations in $\mathcal{H}_{0}^{rep}$.
For instance, certain density models may model the extra configurations in $\mathcal{H}_{0}^{rep}$ more accurately.
However, good performance of the csmGmm across many settings suggests that it may be safer in general.
Additional replication setting details are provided in the Supplementary Materials.
 
 Overall, we see in a variety of settings that the csmGmm is robust in protecting nominal false discovery rates
compared to existing methods, while also preserving reasonable power.
Although the HDMT also generally protects nominal false discovery rates, it can be quite conservative in 
some settings, resulting in considerable power loss.
The DACT, spline-based two-group model, and kernel-based two-group model control false discoveries when mediation
signals are weak and sparse but can become anticonservative in the presence of strong or dense signals, among other situations.
In addition to possessing robust operating characteristics, the csmGmm is also highly flexible and does not produce incongruous results.


\section{Application to Genetic Association Datasets}
 \label{sec:application}
 
\subsection{Mediation}
 \label{ss:mediation}
 
As previously described, our work is motivated by translational genetics studies in 
lung cancer. 
One important analysis that has been difficult to conduct is the search for SNPs that
perturb risk of lung cancer by regulating the expression levels of important genes.
Many studies have implied that this type of mediated effect exists, for example by comparing
lists of SNPs associated with gene expression and lists of gene expressions associated with lung cancer \citep{mckay2017large, bosse2020transcriptome}.
However, formal inference for these effects has been lacking, and few validated results are available.
Identifying the exact SNP-expression-lung cancer pathways could help researchers understand the biological
underpinnings of GWAS association findings or assist with identifying new drug targets.

We focus our investigation on 15 lead risk SNPs identified in the latest ILCCO GWAS, the largest GWAS of lung cancer to date.
Many of these SNPs are hypothesized to affect lung cancer risk by regulating gene expression  \citep{mckay2017large}.
For each SNP, we perform expression quantitative trait loci (eQTL) studies by calculating the association between the SNP and expression of genes in lung tissue
using Genotype-Tissue Expression Project (GTEx) data \citep{lonsdale2013genotype}.
More specifically, for a given SNP, we fit the regression model in Equation (\ref{eq:med1}) 13556 times with GTEx-supplied covariates $\mathbf{X}_{i}$,
varying only the outcome $M_{ij}$ each time to be a different one of the 13556 genes measured in lung tissue.

Next, the ILCCO compendium is used to calculate the association between gene expression values and lung cancer
in an approximation of a recent transcriptome-wide association study \citep{bosse2020transcriptome}.
We focus specifically on subjects with squamous cell lung cancer, utilizing a sample size of 7850 cases and 32319 controls.
The gene expression values for each subject are imputed \citep{barbeira2018exploring} as previously 
described for this dataset \citep{bosse2020transcriptome}.
Then for each combination of the 15 risk SNPs ($G_{i}$ terms) and the 13556 gene expression values ($M_{ij}$ terms), we fit the regression 
model in Equation (\ref{eq:med2}), using lung cancer status as the outcome each time.

We apply the csmGmm separately for each of the 15 SNPs.
Each SNP is involved in 13556 models using Equation (\ref{eq:med1}) and 13556 models using Equation (\ref{eq:med2}).
From these models, we compute the 13556 pairs of test statistics for $(\alpha_{j}, \beta_{j})$
and then apply the csmGmm.
We perform the analysis separately for each SNP because in a pooled analysis, the  Equation (\ref{eq:med2}) test statistics 
for effect of $M_{ij}$ on lung cancer would be almost exactly duplicated 15 times.

The top five findings across all 15 separate analyses are presented below in Table \ref{tab:mediation}. 
These five mediated pathways are the only pairs (out of $15 \times 13556 = 203340$ possible) that are identified
as significant at $q<0.1$ by more than one of the methods considered in simulation;
the csmGmm identified the first, fourth, and fifth rows as significant.
We can see that the SNP rs71658797, which is located near the FUBP1 gene, is identified as a regulator
of the FUBP1, PBX3, and NOTCH4 genes.
Thus it is of interest to prioritize rs71658797, as well as rs6920364 and rs7953330, in further functional mediation studies. 
It would also be of interest to repeat this study with non-imputed lung gene expression data, 
but even large consortiums do not have much access to such data \citep{bosse2020transcriptome}, and imputation has
been shown to provide reliable data in many previous studies \citep{barbeira2018exploring}.

\begin{table}[ht]
\begin{center}
\caption{SNPs with effects on lung cancer risk that are mediated through gene expression in lung tissue. 
The Nearest Gene column shows the gene physically closest to the SNP, while the Mediator column
shows the gene expression that mediates the SNP effect on lung cancer. The numRej column
gives the number of tests (among those considered in simulation) that declared the mediation effect significant at $q<0.1$.}
\label{tab:mediation}
\begin{tabular}{lrrllrrr}
\hline
RS & Chr & BP & Nearest Gene & Mediator & $Z_{\alpha}$ & $Z_{\beta}$ & numRej \\ 
\hline
rs71658797 & 1 & 77967507 & FUBP1 & FUBP1 & -4.26 & -4.12 & 5 \\ 
rs71658797 & 1 & 77967507 & FUBP1 & PBX3 & 2.85 & -4.40 & 2 \\ 
rs71658797 & 1 & 77967507 & FUBP1 & NOTCH4 & 2.65 & -4.18 & 2 \\ 
rs6920364 & 6 & 167376466 & RNASET2 & RNASET2 & -8.70 & 3.22 & 5 \\ 
rs7953330 & 12 & 998819 & RAD52 & RAD52 & -5.85 & 5.71 & 6 \\ 
\hline
\end{tabular}
\end{center}
\end{table}

Note also that the second and third rows of Table \ref{tab:mediation} are the source of the 
incongruous findings example given in Section \ref{sec:intro}.
The 50 degrees of freedom natural spline approach assigns lfdr-values of 0.170 and 0.148 to 
the second and third rows, causing the third row to be rejected while the second is not.
This case succinctly demonstrates the confusion that can be caused by incongruous results.
Over all 203340 composite null sets, the kernel estimator, 50 degrees of freedom spline, and 7 degrees of freedom 
spline produce 10891, 10788, and 8293 incongruous findings, respectively.
 
 \subsection{Pleiotropy}
 \label{ss:pleiotropy}
 
The second primary analysis of interest is a three-way pleiotropy investigation.
Recent promising results in repurposing cardiovascular medications for lung cancer
have spurred much work devoted to characterizing shared
genetic risk factors between lung cancer and related phenotypes such as CAD and
BMI \citep{byun2021shared, pettit2021shared}, 
although few individual pleiotropic SNPs have been identified.
BMI and smoking have been studied together for many years due to the so-called
obesity paradox, where heavier patients appear to demonstrate lower risk for lung cancer.
Data from a GWAS of CAD is available from the CARDIoGRAMplusC4D consortium, with approximately 61000 cases and 123000 controls \citep{nikpay2015comprehensive}. 
Data on BMI is available from the UK Biobank.

We start by using the csmGmm to perform a three-dimensional pleiotropy analysis for lung cancer, CAD, and BMI using
only the ILCCO squamous cell lung cancer GWAS, the CARDIoGRAMplusC4D GWAS, and a UKB GWAS of BMI.
Approximately 6.8 million SNPs are shared among the three datasets, and the number declared significant by 
each method is provided in Table \ref{tab:pleiotropy}.
Interestingly, we can see that the csmGmm finds 0 significant SNPs, while using the other density estimators results in over 200 significant findings. 
This behavior matches the trends observed in simulation, where the other density models do not show reliable error control, especially
in settings with stronger signals.
It is possible to make a case that many significant results are spurious findings attributable to poor false discovery rate control (see Supplementary Materials).

\begin{table}[ht]
\begin{center}
\caption{Number of SNPs found significant in two- and three-way pleiotropy analyses across all methods.
Orig column shows number of significant findings when using nominal $q<0.1$ for all methods, while Adj column shows number of significant findings
when using adjusted nominal thresholds that generate empirical false discovery proportions of 0.1.}
\label{tab:pleiotropy}
\begin{tabular}{lrcrrcrrcrr}
\hline\hline
\multicolumn{1}{c}{\bfseries }&\multicolumn{1}{c}{\bfseries All 3}&\multicolumn{1}{c}{\bfseries }&\multicolumn{2}{c}{\bfseries LC,CAD}&\multicolumn{1}{c}{\bfseries }&\multicolumn{2}{c}{\bfseries LC,BMI}&\multicolumn{1}{c}{\bfseries }&\multicolumn{2}{c}{\bfseries CAD,BMI}\tabularnewline
\cline{2-2} \cline{4-5} \cline{7-8} \cline{10-11}
\multicolumn{1}{c}{Method}&\multicolumn{1}{c}{}&\multicolumn{1}{c}{}&\multicolumn{1}{c}{Orig}&\multicolumn{1}{c}{Adj}&\multicolumn{1}{c}{}&\multicolumn{1}{c}{Orig}&\multicolumn{1}{c}{Adj}&\multicolumn{1}{c}{}&\multicolumn{1}{c}{Orig}&\multicolumn{1}{c}{Adj}\tabularnewline
\hline
csmGmm &$  0$&&$315$&$315$&&$4079$&$4079$&&$1713$&$1713$\tabularnewline
Kernel&$239$&&$710$&$305$&&$9001$&$3489$&&$7898$&$1380$\tabularnewline
locfdr50&$246$&&$713$&$306$&&$8966$&$3486$&&$7966$&$1382$\tabularnewline
locfdr7&$216$&&$714$&$300$&&$9554$&$3474$&&$7567$&$1152$\tabularnewline
HDMT&$-$&&$283$&$283$&&$2459$&$2459$&&$915$&$ 915$\tabularnewline
DACT&$-$&&$345$&$234$&&$9234$&$3032$&&$8869$&$2419$\tabularnewline
\hline
\end{tabular}
\end{center}
\end{table}

Given the lack of csmGmm significant findings in the three-way analysis, we additionally test the three two-way pleiotropy composite null hypotheses as well. 
The raw, unadjusted findings are given in Table \ref{tab:pleiotropy}, in the Orig column.
We can see that again the csmGmm finds far fewer significant associations.
However, the comparison is not very fair
because we observed in simulation that some other methods do not provide reliable error control, especially in settings
where large datasets produce stronger signals.
To create a more equitable comparison, we perform simulations (see Supplementary Materials for details) using the analysis data
to estimate nominal error rates that would result in similar empirical false discovery proportions for each method.
The number of findings using adjusted error rates are reported in the Adj column, except for the csmGmm and HDMT, which are unadjusted.
After adjustment, we can see that most methods identify similar numbers of significant SNPs.
Again, this pattern broadly follows the trends observed in simulation.

\begin{figure}[!ht]
\begin{center}
\centerline{\includegraphics[scale=0.5, angle=0]{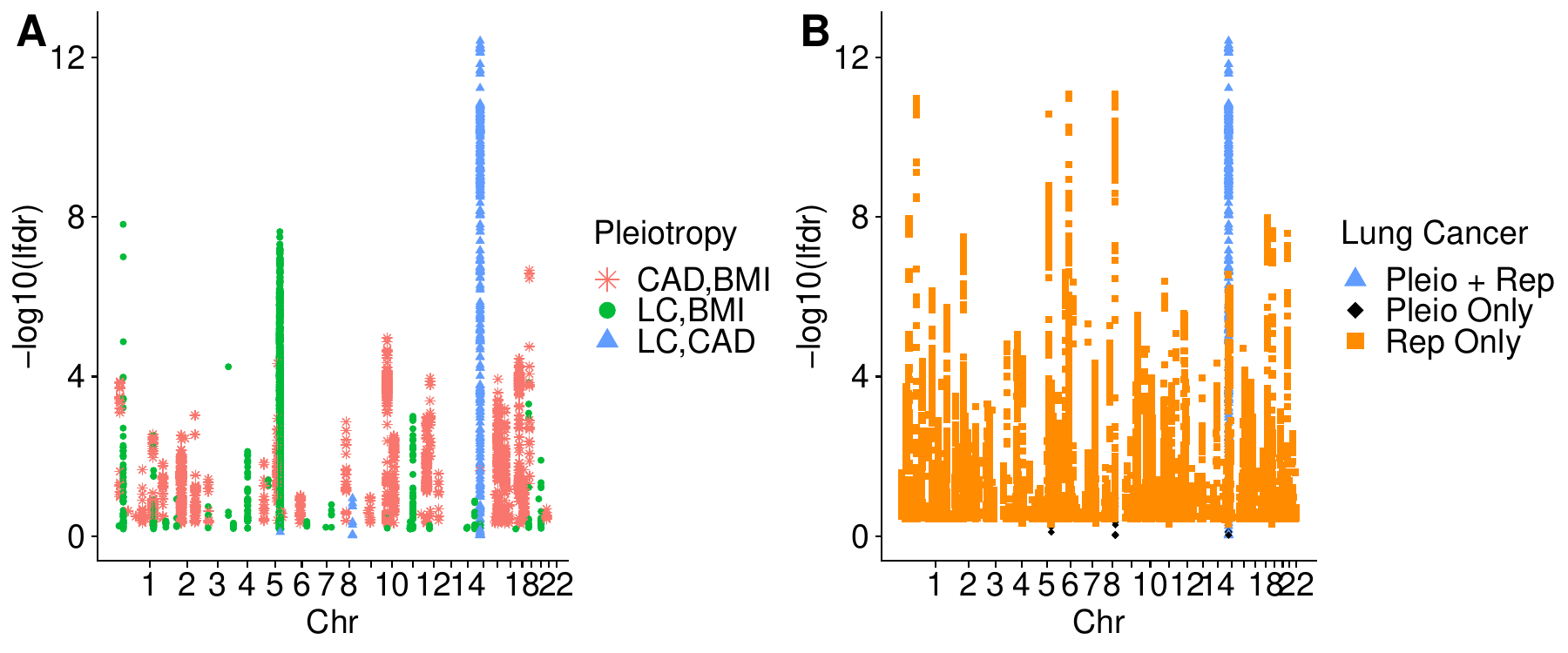}}
\end{center}
\caption{
Manhattan plots of pleiotropy analysis. Panel A shows significant two-way pleiotropy findings according to the csmGmm. 
In B, blue triangles (Pleio + Rep) indicate the SNP is identified in all three of lung cancer-CAD pleiotropy analysis 
(using ILCCO and CARDIoGRAMplusC4D), lung cancer replication analysis (using ILCCO and UKB), and
CAD replication analysis (using CARDIoGRAMplusC4D and UKB).
Black diamonds (Pleio Only) indicate the SNP is identified in lung cancer-CAD pleiotropy analysis
but not both replication analyses. 
Orange squares indicate the SNP is identified in at least one replication analysis but not lung cancer-CAD pleiotropy analysis.
Y-axis gives pleiotropy lfdr-value.}
\label{fig:manhattans}
\end{figure}

For more insight into the shared genetic architectures of lung cancer, CAD, and BMI, we visualize
the csmGmm findings in Figure \ref{fig:manhattans}.
Figure \ref{fig:manhattans}A shows a plot of the lfdr-values for each of the two-way analyses.
Many top lung cancer-CAD SNPs fall on chromosome 15 near the gene ADAMTS7.
This region has been discussed extensively as a risk locus for CAD and other
inflammation-related diseases like squamous cell lung cancer \citep{arroyo2015adamts7}.
It also lies very near the CHRNA3 and CHRNA5 nicotine receptor genes, which are well-known to demonstrate
strong associations with lung cancer \citep{mckay2017large}. 
For lung cancer and BMI, many significant findings lie near the major histocompatibility complex on
chromosome 6, again a region that has been implicated numerous times in inflammatory diseases \citep{mckay2017large}. 
Taken together, these results all provide reassurance that the csmGmm is identifying true variants of 
interest.

In Figure \ref{fig:manhattans}B we overlap lung cancer-CAD pleiotropy findings with within-lung cancer and within-CAD replication analyses.
The UKB is used to replicate both ILCCO and CARDIoGRAMplusC4D.
SNPs that are pleiotropic for lung cancer and CAD, replicated for lung cancer, and replicated for CAD are colored in blue.
We see that most of the top pleiotropic SNPs near nicotine receptor genes are indeed also replicated findings for
both lung cancer and CAD.
Also, the number of SNPs identified as pleiotropic that are not replicated is very small, suggesting 
that the pleiotropy analysis is truly identifying important variants.
A UKB-only pleiotropy analysis with the c-csmGmm is available in the
Supplementary Materials.

\section{Discussion}
 \label{sec:discussion}
 
We have proposed the conditionally symmetric multidimensional Gaussian mixture model framework to
unify and generalize many common genetic analyses that require inference on composite null hypotheses.
A key theme of the csmGmm density model is to exploit the similarities of test statistics generated from the same binary representation; 
all association configurations with null hypotheses in the same dimensions share a binary representation.
Conditional on the binary representation, each dimension of a test statistic vector is assumed to be generated from a symmetric
mixture of Gaussian distributions.
The two-group approach and EM algorithm can then be used to estimate unknown parameters and calculate local false discovery rates, which
are used for inference.

The base csmGmm is highly flexible and can be used to perform mediation studies or pleiotropy
studies of any dimension when the outcomes are taken from independent cohorts.
We further introduce the c-csmGmm for composite null sets with correlated test statistics,
which are found when pleiotropy studies are conducted on outcomes measured in the 
same cohort.
Finally, we propose the r-csmGmm for replication studies, which require that effects possess the same direction under
the alternative and thus utilize the more complex composite null $H_{0}^{rep}$.
The software to perform all tests described in this work is publicly available in the R package \verb|csmGmm|.

A critical differentiating feature of the csmGmm is its ability to produce interpretable empirical Bayes inference that does not contradict
frequentist  z-statistic multiple testing rules.
In general, contradictions between lfdr-values and z-statistics can arise easily and often, as demonstrated in our motivating applications.
We prove that all three of the csmGmm, c-csmGmm, and r-csmGmm prevent incongruous results in practically important
testing settings.
The harmonization of lfdr-value and z-statistic testing rules is especially relevant for genetic association studies, 
which rely heavily on summary statistics.

Extensive empirical evidence demonstrates that the csmGmm offers robust control of false discoveries
while maintaining power comparable to other methods, outperforming recently-developed
methodology in many realistic genetic association situations.
While other strategies perform well in a subset of simulations, only the csmGmm is able to protect nominal
false discovery rates across a wide range of situations while not sacrificing power.
The HDMT also offers robust protection of false discovery rates, but its conservative nature
results in lower power.

The DACT and two-group approaches with other density models control the nominal
error rate when signals are weak and sparse in mediation settings.
However, the DACT can be anticonservative when effect sizes are large
or when the proportion of non-zero effects is too high.
The other two-group approaches may additionally struggle in higher-dimensional settings when the proportion of non-zero effects is too low or 
effect sizes are large. 
As genetic sample sizes continue to increase and more massive compendiums are planned \citep{bycroft2018uk},
it will be more and more important for composite null tools to offer reliable performance when associations
are stronger and larger composite null sets are studied.
Additionally, the DACT and HDMT are not available for settings with more than two dimensions or replication studies.
Only the c-csmGmm is designed for sets with correlated test statistics.

Application to a variety of lung cancer investigations integrating disparate datasets demonstrates
that the csmGmm is able to recapitulate known findings about lung cancer and related diseases.
Specifically, mediation analysis shows evidence of SNP effects on lung cancer that are mediated through
expression of FUBP1, RNASET2, and RAD52;
the latter two genes were strongly implicated in previous publications \citep{mckay2017large}. 
We also find evidence of pleiotropic SNPs associated with both lung cancer and CAD at known risk loci on chromosome 15.
Most of the pleiotropic SNPs can be replicated both within lung cancer and within CAD.
Interestingly, the csmGmm suggests the lack of genetic variants with three-way pleiotropic effects on lung cancer, CAD, and BMI.

It is of future research interest to determine how conditions in the csmGmm can be  
relaxed while retaining good performance and interpretability.
For example, it is possible the normality constraints could be loosened to shape constraints.
Such modifications would provide more flexibility for use on a wider variety of datasets.
It would additionally be of interest to propose models for larger sets of correlated test statistics.
Currently the c-csmGmm is only developed for two-dimensional sets, but there exist situations
where it would be advantageous to consider more correlated test statistics.
It is not clear how to prove a version of Theorem 2 for sets with $K>2$ or arbitrary correlation structures.

\begin{center}
{\bf SUPPLEMENTARY MATERIAL}
\end{center}

\noindent Supplementary Materials provide the proofs of Theorems 1-3, provide further details about the EM algorithm implementation, and expand on the simulation and data analysis results.

\bibliographystyle{apalike}
\bibliography{empBayesBib}

\begin{thebibliography}{}

\bibitem[Arroyo and Andr{\'e}s, 2015]{arroyo2015adamts7}
Arroyo, A.~G. and Andr{\'e}s, V. (2015).
\newblock {ADAMTS7} in cardiovascular disease: from bedside to bench and back
  again?
\newblock {\em Circulation}, 131(13):1156--1159.

\bibitem[Barbeira et~al., 2018]{barbeira2018exploring}
Barbeira, A.~N., Dickinson, S.~P., Bonazzola, R., Zheng, J., Wheeler, H.~E.,
  Torres, J.~M., Torstenson, E.~S., Shah, K.~P., Garcia, T., Edwards, T.~L.,
  et~al. (2018).
\newblock Exploring the phenotypic consequences of tissue specific gene
  expression variation inferred from {GWAS} summary statistics.
\newblock {\em Nature Communications}, 9(1):1--20.

\bibitem[Barfield et~al., 2017]{barfield2017testing}
Barfield, R., Shen, J., Just, A.~C., Vokonas, P.~S., Schwartz, J., Baccarelli,
  A.~A., VanderWeele, T.~J., and Lin, X. (2017).
\newblock Testing for the indirect effect under the null for genome-wide
  mediation analyses.
\newblock {\em Genetic Epidemiology}, 41(8):824--833.

\bibitem[Benjamini and Hochberg, 1995]{benjamini1995controlling}
Benjamini, Y. and Hochberg, Y. (1995).
\newblock Controlling the false discovery rate: a practical and powerful
  approach to multiple testing.
\newblock {\em Journal of the Royal Statistical Society: Series B
  (Methodological)}, 57(1):289--300.

\bibitem[Boss{\'e} et~al., 2020]{bosse2020transcriptome}
Boss{\'e}, Y., Li, Z., Xia, J., Manem, V., Carreras-Torres, R., Gabriel, A.,
  Gaudreault, N., Albanes, D., Aldrich, M.~C., Andrew, A., et~al. (2020).
\newblock Transcriptome-wide association study reveals candidate causal genes
  for lung cancer.
\newblock {\em International Journal of Cancer}, 146(7):1862--1878.

\bibitem[Bycroft et~al., 2018]{bycroft2018uk}
Bycroft, C., Freeman, C., Petkova, D., Band, G., Elliott, L.~T., Sharp, K.,
  Motyer, A., Vukcevic, D., Delaneau, O., O’Connell, J., et~al. (2018).
\newblock The {UK Biobank} resource with deep phenotyping and genomic data.
\newblock {\em Nature}, 562(7726):203--209.

\bibitem[Byun et~al., 2022]{byun2022cross}
Byun, J., Han, Y., Li, Y., Xia, J., Long, E., Choi, J., Xiao, X., Zhu, M.,
  Zhou, W., Sun, R., et~al. (2022).
\newblock Cross-ancestry genome-wide meta-analysis of 61,047 cases and 947,237
  controls identifies new susceptibility loci contributing to lung cancer.
\newblock {\em Nature Genetics}, 54(8):1167--1177.

\bibitem[Byun et~al., 2021]{byun2021shared}
Byun, J., Han, Y., Ostrom, Q.~T., Edelson, J., Walsh, K.~M., Pettit, R.~W.,
  Bondy, M.~L., Hung, R.~J., McKay, J.~D., and Amos, C.~I. (2021).
\newblock The shared genetic architectures between lung cancer and multiple
  polygenic phenotypes in genome-wide association studies.
\newblock {\em Cancer Epidemiology and Prevention Biomarkers},
  30(6):1156--1164.

\bibitem[Chen and Li, 2009]{chen2009hypothesis}
Chen, J. and Li, P. (2009).
\newblock Hypothesis test for normal mixture models: The {EM} approach.
\newblock {\em The Annals of Statistics}, 37(5A):2523--2542.

\bibitem[Dai et~al., 2022]{dai2022multiple}
Dai, J.~Y., Stanford, J.~L., and LeBlanc, M. (2022).
\newblock A multiple-testing procedure for high-dimensional mediation
  hypotheses.
\newblock {\em Journal of the American Statistical Association},
  117(537):198--213.

\bibitem[Dempster et~al., 1977]{dempster1977maximum}
Dempster, A.~P., Laird, N.~M., and Rubin, D.~B. (1977).
\newblock Maximum likelihood from incomplete data via the {EM} algorithm.
\newblock {\em Journal of the Royal Statistical Society: Series B
  (Methodological)}, 39(1):1--22.

\bibitem[Efron, 2008]{efron2008microarrays}
Efron, B. (2008).
\newblock Microarrays, empirical bayes and the two-groups model.
\newblock {\em Statistical Science}, 23(1):1--22.

\bibitem[Efron and Tibshirani, 2002]{efron2002empirical}
Efron, B. and Tibshirani, R. (2002).
\newblock Empirical bayes methods and false discovery rates for microarrays.
\newblock {\em Genetic Epidemiology}, 23(1):70--86.

\bibitem[Heller and Yekutieli, 2014]{heller2014replicability}
Heller, R. and Yekutieli, D. (2014).
\newblock Replicability analysis for genome-wide association studies.
\newblock {\em The Annals of Applied Statistics}, 8(1):481--498.

\bibitem[Huang, 2019]{huang2019genome}
Huang, Y.-T. (2019).
\newblock Genome-wide analyses of sparse mediation effects under composite null
  hypotheses.
\newblock {\em The Annals of Applied Statistics}, 13(1):60--84.

\bibitem[Lindsay, 1988]{lindsay1988composite}
Lindsay, B.~G. (1988).
\newblock Composite likelihood methods.
\newblock {\em Contemporary Mathematics}, 80(1):221--239.

\bibitem[Liu and Lin, 2018]{liu2018multiple}
Liu, Z. and Lin, X. (2018).
\newblock Multiple phenotype association tests using summary statistics in
  genome-wide association studies.
\newblock {\em Biometrics}, 74(1):165--175.

\bibitem[Liu et~al., 2022]{liu2022large}
Liu, Z., Shen, J., Barfield, R., Schwartz, J., Baccarelli, A.~A., and Lin, X.
  (2022).
\newblock Large-scale hypothesis testing for causal mediation effects with
  applications in genome-wide epigenetic studies.
\newblock {\em Journal of the American Statistical Association},
  117(537):67--81.

\bibitem[Lonsdale et~al., 2013]{lonsdale2013genotype}
Lonsdale, J., Thomas, J., Salvatore, M., Phillips, R., Lo, E., Shad, S., Hasz,
  R., Walters, G., Garcia, F., Young, N., et~al. (2013).
\newblock The genotype-tissue expression {(GTEx)} project.
\newblock {\em Nature Genetics}, 45(6):580--585.

\bibitem[McKay et~al., 2017]{mckay2017large}
McKay, J.~D., Hung, R.~J., Han, Y., Zong, X., Carreras-Torres, R., Christiani,
  D.~C., Caporaso, N.~E., Johansson, M., Xiao, X., Li, Y., et~al. (2017).
\newblock Large-scale association analysis identifies new lung cancer
  susceptibility loci and heterogeneity in genetic susceptibility across
  histological subtypes.
\newblock {\em Nature Genetics}, 49(7):1126--1132.

\bibitem[Nikpay et~al., 2015]{nikpay2015comprehensive}
Nikpay, M., Goel, A., Won, H.-H., Hall, L.~M., Willenborg, C., Kanoni, S.,
  Saleheen, D., Kyriakou, T., Nelson, C.~P.,  et~al. (2015).
\newblock A comprehensive 1000 genomes-based genome-wide association
  meta-analysis of coronary artery disease.
\newblock {\em Nature Genetics}, 47(10):1121.

\bibitem[Pettit et~al., 2021]{pettit2021shared}
Pettit, R.~W., Byun, J., Han, Y., Ostrom, Q.~T., Edelson, J.,
  Bondy, M.~L., Hung, R.~J., McKay, J.~D., and Amos, C.~I. (2021).
\newblock The shared genetic architecture between epidemiological and
  behavioral traits with lung cancer.
\newblock {\em Scientific Reports}, 11(1):1--12.

\bibitem[Stephens, 2017]{stephens2017false}
Stephens, M. (2017).
\newblock False discovery rates: a new deal.
\newblock {\em Biostatistics}, 18(2):275--294.

\bibitem[Sun and Cai, 2007]{sun2007oracle}
Sun, W. and Cai, T.~T. (2007).
\newblock Oracle and adaptive compound decision rules for false discovery rate
  control.
\newblock {\em Journal of the American Statistical Association},
  102(479):901--912.

\bibitem[Visscher et~al., 2017]{visscher201710}
Visscher, P.~M., Wray, N.~R., Zhang, Q., Sklar, P., McCarthy, M.~I., Brown,
  M.~A., and Yang, J. (2017).
\newblock 10 years of gwas discovery: biology, function, and translation.
\newblock {\em The American Journal of Human Genetics}, 101(1):5--22.

\bibitem[Wang and Wei, 2020]{wang2020imix}
Wang, Z. and Wei, P. (2020).
\newblock Imix: a multivariate mixture model approach to association analysis
  through multi-omics data integration.
\newblock {\em Bioinformatics}, 36(22-23):5439--5447.

\bibitem[Willer et~al., 2013]{willer2013discovery}
Willer, C.~J., Schmidt, E.~M., Sengupta, S., Peloso, G.~M., Gustafsson, S.,
  Kanoni, S., Ganna, A., Chen, J., Buchkovich, M.~L., Mora, S., et~al. (2013).
\newblock Discovery and refinement of loci associated with lipid levels.
\newblock {\em Nature Genetics}, 45(11):1274.

\end{thebibliography}

\end{document}